\documentclass[11pt,a4paper]{article}

\RequirePackage{ifpdf} 
\usepackage{amsmath} 
\usepackage{mathtools}

\usepackage{tikz}
\usepackage[compat=1.1.0]{tikz-feynman}
\usetikzlibrary{decorations.pathmorphing}
\usetikzlibrary{positioning,arrows}
\usetikzlibrary{decorations.markings}
\usetikzlibrary{shapes,shadows}

\usepackage{jheppub}
\usepackage{pstricks}
\usepackage[final]{pdfpages} 
\usepackage{ifpdf} 
\usepackage{slashed}
\usepackage{hyperref}

\usepackage{color} 
\usepackage{graphics}
\usepackage{float}

\usepackage{etoolbox} 
\usepackage{fixmath}
\usepackage{mathrsfs}

\usepackage{amsfonts}

\usepackage{epstopdf}

\usepackage[normalem]{ulem}
\usepackage{framed}
\usepackage{comment}
\usepackage{longtable}
\usepackage{booktabs}

\newcommand{\ep}{\epsilon}
\newcommand{\I}{{\cal I}}
\newcommand{\II}{{\cal I}}

\newcommand{\pp}{{\cal P}}

\title{Three loop master integrals for ${\mathcal{O}} (\alpha \alpha_s^2)$ corrections to quark form factor}

\author[a]{Tanmoy Pati,}
\author[a]{Narayan Rana}

\affiliation[a]{School of Physical Sciences, National Institute of Science Education and Research,\\ An OCC of Homi Bhabha National Institute, Jatni 752050, India}

\emailAdd{tanmoy.pati@niser.ac.in, narayan.rana@niser.ac.in}

\abstract{
We consider the three-loop mixed strong-electroweak (${\mathcal{O}}(\alpha \alpha_s^2)$) corrections to the quark form factor. We compute the master integrals which are appearing in the Feynman diagrams containing a single massive boson in the loop. We use the state-of-the-art method of differential equations to compute all 303 of them, expressing the results in terms of generalized polylogarithms. 
We encounter multiple square roots that cannot be simultaneously rationalized using a single transformation. Applying concurrent transformations allows us to express the results through generalized polylogarithms with a simple alphabet, but with multiple interdependent arguments.
}

\colorlet{shadecolor}{gray!14}

\begin{document}

\tikzset{
photon/.style={decorate, draw=black,
    decoration={coil,aspect=0,segment length=7pt,amplitude=2pt}},    
Zboson/.style={decorate, draw=red,
    decoration={snake,aspect=0,segment length=5pt,amplitude=2pt}},    
gluon/.style={decorate, draw=black,
    decoration={coil,aspect=0.5,segment length=5pt,amplitude=1pt}},
fermion/.style={draw=black,
      postaction={decorate},decoration={markings,mark=at position .55
        with {\arrow[draw=black]{>}}}},   
scalar/.style={draw=black,
      postaction={decorate},decoration={}},           
vector/.style={decorate, decoration={snake}, draw} 
}

\preprint{~}

\allowdisplaybreaks[4]
\unitlength1cm
\maketitle
\flushbottom


\section{Introduction}

Scattering amplitudes are essential for computing hard scattering processes in perturbative Quantum Chromodynamics (QCD). Beyond delivering precise phenomenological predictions, the scattering
amplitudes provide a clear insight into the underlying principles of Quantum Field Theory such as factorization or the universality of infrared (IR) singularities. 
The simplest amplitudes, known as form factors, involve two on-shell states of elementary fields,
either both massless (quarks or gluons)~\cite{Ravindran:2004mb,deFlorian:2013sza,Moch:2005tm,Moch:2005id,Baikov:2009bg,Gehrmann:2010ue,Gehrmann:2014vha,Ahmed:2015qia,Ahmed:2015qpa,Ahmed:2016vgl,Ahmed:2016qjf,Ahmed:2019yjt,Lee:2021lkc,Lee:2022nhh,Chakraborty:2022yan}
or both massive (quarks)~\cite{Bernreuther:2004ih,Bernreuther:2004th,Bernreuther:2005rw,Bernreuther:2005gw,Gluza:2009yy,Ablinger:2017hst,Henn:2016tyf,Lee:2018nxa,Ablinger:2018yae,Lee:2018rgs,Blumlein:2019oas,Fael:2022rgm,Fael:2022miw,Fael:2023zqr,Blumlein:2023uuq}
or one massless and one massive~\cite{Bonciani:2008wf,Huber:2009se,Bell:2006tz,Bell:2010mg,Chen:2018dpt,Engel:2018fsb,Datta:2023otd,Fael:2024vko,Datta:2024cen},
and an off-shell state described through a composite operator. 
One such form factor, the quark form factor, plays a significant role in precise phenomenological predictions for the Drell-Yan (DY) production of a lepton pairs.
This process stands as a cornerstone for physics investigations at the Large Hadron Collider (LHC). Beyond its role in precisely determining key parameters of the weak interaction, such as the sine of the weak mixing angle and the W boson mass, the DY process is instrumental for constraining parton distribution functions (PDFs), calibrating detectors, and establishing collider luminosity. Furthermore, its final state signatures closely resemble those predicted by numerous beyond-the-Standard-Model (BSM) theories, making it a critical Standard Model (SM) background in the search for New Physics.

Given its significant relevance, the DY process has been measured experimentally with great 
precision and has also been computed theoretically to a high degree of accuracy. 
Indeed, the DY process was one of the earliest to have radiative corrections computed, considering both strong ($\alpha_s$) and electroweak (EW) ($\alpha$) couplings. The next-to-leading-order (NLO)~\cite{Altarelli:1979ub} and next-to-next-to-leading-order (NNLO)~\cite{Hamberg:1990np,Harlander:2002wh} QCD corrections to the total cross section were followed by differential NNLO calculations incorporating leptonic decays~\cite{Anastasiou:2003yy,Anastasiou:2003ds,Melnikov:2006kv,Catani:2009sm,Catani:2010en}. Complete EW corrections have been calculated for $W$~\cite{Dittmaier:2001ay,Baur:2004ig,Zykunov:2006yb,Arbuzov:2005dd,CarloniCalame:2006zq} and $Z$~\cite{Baur:2001ze,Zykunov:2005tc,CarloniCalame:2007cd,Arbuzov:2007db,Dittmaier:2009cr} production. Recent advancements include next-to-next-to-next-to-leading-order (N$^3$LO) QCD radiative calculations 
for both the inclusive~\cite{Duhr:2020seh,Chen:2021vtu,Duhr:2020sdp}
as well as fiducial~\cite{Camarda:2021ict,Chen:2022cgv,Neumann:2022lft,Campbell:2023lcy}
production cross section,
and NNLO mixed QCD-EW corrections~\cite{Dittmaier:2014qza,Dittmaier:2015rxo,Bonciani:2016wya,Bonciani:2019nuy,Bonciani:2020tvf,Buccioni:2020cfi,Behring:2020cqi,Buonocore:2021rxx,Bonciani:2021zzf,Bonciani:2021iis,Buccioni:2022kgy,Armadillo:2022bgm,Dittmaier:2024row,Armadillo:2024nwk}.
Notably, these mixed QCD-EW corrections have proven larger than anticipated, highlighting the 
necessity of including other corrections such as N$^3$LO mixed 
QCD-EW (${\mathcal{O}} (\alpha \alpha_s^2)$) corrections.
Quark form factors at ${\mathcal{O}} (\alpha \alpha_s^2)$
are a key component in obtaining these higher-order corrections.

The state-of-the-art approach to computing these virtual amplitudes involves reducing scalar Feynman integrals to a linearly independent basis, known as Master Integrals (MIs), through 
integration-by-parts (IBP)~\cite{Tkachov:1981wb,Chetyrkin:1981qh,Laporta:2001dd} and 
Lorentz invariance (LI) identities. Subsequently, the MIs are computed using the method of differential equations~\cite{Kotikov:1990kg,Remiddi:1997ny,Gehrmann:1999as,Argeri:2007up,Henn:2013pwa,Henn:2014qga,Ablinger:2015tua,Ablinger:2018zwz}.
The basic principle of this technique involves differentiating the MIs with respect to the kinematic variables and then applying IBP identities to the resulting expressions. This process yields a system of first-order coupled differential equations. By strategically organizing this system into a block-triangular form, a solution can be obtained either through a bottom-up or top-down approach.
The sub-systems can be decoupled to form higher-order differential equations, which are subsequently solved using the method of variation of constants.
This powerful technique can even be improved \cite{Henn:2013pwa,Henn:2014qga} if the system can be reduced to canonical form or $\ep$-form 
where each MIs can be solved in terms of 
iterated integrals such as harmonic polylogarithms (HPLs)~\cite{Remiddi:1999ew},
generalized harmonic polylogarithms (GPLs)~\cite{Goncharov:2001iea,Vollinga:2004sn},
Chen iterated integrals~\cite{Chen:1977oja} etc.
In this paper, we present the computation of MIs relevant to the ${\mathcal{O}} (\alpha \alpha_s^2)$ corrections to quark form factors at three loops. We categorize the contributing Feynman diagram topologies into three groups based on the EW bosons involved. First, diagrams with a photon in the loop, photon being massless, exhibit topologies that are subsets of those found in three-loop QCD corrections. Second, diagrams featuring a triple vector boson vertex have topologies that are again subsets, this time of those appearing in three-loop mixed QCD-EW corrections to Higgs boson production, 
as presented in \cite{Bonetti:2017ovy}. Third, we consider Feynman diagrams containing a single $Z$ or $W$ boson. The MIs appearing in these topologies are novel and are the focus of this paper.

The remainder of this paper is organized as follows. Section \ref{sec:notation} establishes our notation and details the choice of the Feynman prescription for all introduced variables. In Section \ref{sec:comp}, we detail the computational approach, specifying the integral families, presenting our choice of MIs, and commenting on their analytic evaluation using the method of differential equations. The findings of this work are presented in Section \ref{sec:result}, and Section \ref{sec:conclu} provides the concluding remarks.


\section{Notation}
\label{sec:notation}
In this section, we establish our notation. 
We define the physical scattering process, and introduce the required dimensionless variables. 
We consider the scattering process of an off-shell vector boson ($Z$) production with virtuality $q^2$,
in quark-antiquark annihilation
\begin{equation}
    q (p_1) + \bar{q} (p_2) \rightarrow Z (q) \,,
\end{equation}
where the incoming quark and anti-quark carry $p_1$ and $p_2$ momenta, respectively,
with the on-shell conditions 
\begin{equation}
    p_1^2 = 0 \,, ~~ p_2^2 = 0\,.
\end{equation}
We also introduce the following dimensionless variables
\begin{equation}\label{eq1}
    -\frac{q^2}{m_V^2} = -\frac{s}{m_V^2} = x = \frac{(1+x_l)^2}{x_l} = \frac{x_n}{(1+x_n)^2} = -x_i^2 \,,
\end{equation}
where $m_V$ denotes the mass of the vector boson ($Z$ or $W$) present in the loop.
We consider the three-loop mixed QCD-EW corrections (${\mathcal{O}}(\alpha \alpha_s^2)$) 
to this reaction.
As mentioned earlier, we focus on the MIs linked to topologies in Feynman diagrams 
that include a massive vector boson propagator in the loop.

We have introduced three parameters $x_l$, $x_n$ and $x_i$ as suitable base transformations to rationalize the roots that
appear while solving the differential equations. 
Therefore, the final expressions for our results involve GPLs with arguments $x, x_l, x_n$ and $x_i$, 
as well as polynomials of these variables. 
The variable $x$ is defined such that, in the unphysical region ($s < 0$), x is real and positive.
To perform the analytic continuation to the physical region, the Feynman prescription
on the invariants is required. In the physical region, $s$ is positive with a 
positive infinitesimal imaginary part, $s+i0^+$.
Hence, $x$ is given by
\begin{equation}
    x \rightarrow x - i 0^+ \,.
\end{equation}
In terms of $s$, $x_l$ can be solved to obtain the following two roots
\begin{equation}
     x_l^{(\pm)} =  \frac{\sqrt{s + i 0^+}\mp\sqrt{4 m_V^2 + s + i 0^+}}{\sqrt{s + i 0^+}\pm\sqrt{4 m_V^2 + s + i 0^+}} \,.
\end{equation}
\begin{enumerate}
    \item For $s \geq 0$, both $x_l^{(\pm)}$ are negative-valued real numbers, with modulus less than or greater than one, respectively. The prescription becomes $x_l^{(\pm)} = -w_{\pm} \mp i0^+$, where $w_{\pm}$ are the absolute values of the roots.
    \item For $-4 m_V^2 < s < 0$, both $x_l^{(\pm)}$ are complex numbers that lie on the 
    upper and lower half of the unit circle, respectively.
    \item For $s \leq -4m_V^2$, both $x_l^{(\pm)}$ are positive-valued real numbers, with modulus less than and greater than one, respectively.
\end{enumerate}
\begin{figure}[h]
    \centering
    \includegraphics[scale=1.4]{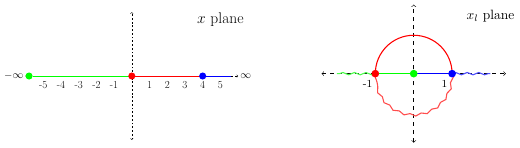}
    \caption{
    The figure illustrates the variable transformation between $x$ and $x_l$. 
    The left and right panel displays the $x$-plane and 
    the complex $x_l$-plane, respectively. 
    Colored lines show the mapping of intervals. In the $x_l$-plane, 
    straight and wiggly lines represent two distinct roots.
    }
    \label{fig:x-xl}
\end{figure}
In Fig.\ref{fig:x-xl}, we illustrate the transformation $x \rightarrow x_l$.
We choose the root $x_l^{(+)}$ which maps the real axis of the complex $x$-plane into the unit circle (solid lines)
in the $x_l$-plane.
The kinematic points $x \rightarrow 0, x \rightarrow 4$ and $x \rightarrow \pm \infty$ in the $x$-plane correspond to 
$x_l \rightarrow -1$ (red dot), $x_l \rightarrow 1$ (blue dot) and $x_l \rightarrow 0^{\pm}$ (green dot), respectively. 
The intervals $x < 0$, $0 < x < 4$ and $x > 4$ are mapped to 
$-1 < x_l < 0$, the upper semi-circle, and $0 < x_l < 1$, respectively.
Due to our chosen Feynman prescription ($s+i0^+$), the green and blue lines in the $x_l$ plane lie infinitesimally 
above and below the real axis, respectively.

Similarly, $x_n$ can be solved to obtain the following two roots. 
\begin{equation}
    x_n(\pm)=\frac{m_V \mp \sqrt{m_V^2 + 4 s + i 0^+}}{m_V \pm \sqrt{m_V^2 + 4 s + i 0^+}} \,.
\end{equation}
\begin{enumerate}
    \item For $s > 0$, both roots $x_n^{(\pm)}$ are negative-valued real numbers, with modulus less than or greater than one, respectively.
    Our choice of the Feynman prescription gives $x_n^{(\pm)} = -{w_n}_\pm \pm i 0^+$, where $w_n$ are the moduli of the roots. 
    \item For $-\frac{m_V^2}{4} < s < 0$, the roots $x_n^{(\pm)}$ are positive-valued real numbers, with modulus less than or greater than one, respectively. 
    \item For $s \leq -\frac{m_V^2}{4}$, the roots $x_n^{(\pm)}$ are complex numbers that lie on the 
    lower and upper half of the unit circle, respectively.
\end{enumerate}
\begin{figure}[h]
    \centering
    \includegraphics[scale=1.4]{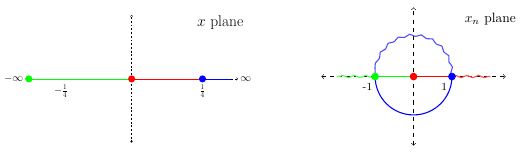}
    \caption{Same as Fig.~\ref{fig:x-xl}, but for the variable transformation between $x$ and $x_n$.}
    \label{fig:x-xn}
\end{figure}
In Fig.\ref{fig:x-xn}, we illustrate the transformation $x \rightarrow x_n$.
We choose the root $x_n^{(+)}$ which maps the real axis of the complex $x$-plane into the unit circle (solid lines)
in the $x_n$-plane.

$x_i$ can be trivially solved in terms of $s$ as
\begin{equation}
    x_i^{(\pm)} = \mp \sqrt{\frac{s+i0^+}{m_V^2}} \,.
\end{equation}
In the physical region, both roots are real. Conversely, Feynman integrals are complex, so expressing them using GPLs in terms of $x_i$ introduces an explicit $i \pi$ term. The sign of this term inherently depends on the prescription chosen.
To present prescription-independent results, we explicitly isolate the logarithms 
from all GPLs and consistently express logarithms of $x_i$ to $\log(x)$.


\section{Computational details}
\label{sec:comp}
We adopt the conventional approach for obtaining and computing the MIs. 
Initially, we have identified the integral families and their corresponding sectors that
fully encompass the relevant Feynman diagrams which appear in the physical process. 
The identification has been performed using \textsc{Reduze}~\cite{vonManteuffel:2012np}.
We have then carried out the IBP 
reduction with the help of \textsc{Kira}~\cite{Maierhofer:2017gsa,Klappert:2020nbg}, which yielded a total of 303 MIs. Subsequently, 
we have utilized the method of differential equations to solve these MIs. 
The specifics of each step are detailed in this section.
\subsection{Integral families}
Each integral family contains three auxiliary propagators.
Thus, a strategic selection would result in a minimal set of integral families. 
However, for convenience in the IBP reduction process, we deliberately choose the
following 25 integral families.
We begin by defining the notation to represent the inverse of the propagator within these integral families as follows.
\begin{align}
    \pp_{i} = k_i^2 \,, ~~~
    \pp_{ij} = (k_i-k_j)^2 \,, ~~~
    \pp_{123} = (k_1+k_2-k_3)^2 \,, ~~~
    \pp_{i;j} = (k_i-p_j)^2 \,, ~~
\nonumber\\
    \pp_{i;12} = (k_i-p_1-p_2)^2 \,, ~~
    \pp_{23;1} = (k_2-k_3-p_1)^2 \,, ~~
    \pp_{23;-2} = (k_2-k_3+p_2)^2 \,.
\end{align}
In the preceding definitions, the indices $i$ and $j$ take the values 1, 2, and 3. 
Now we present the 25 integral families.
\begin{align*}
& \I_{1} : \{
\pp_{1}, \pp_{2}, \pp_{3}, \pp_{12}-m_V^2, \pp_{23}, \pp_{13}, \pp_{1;1}, \pp_{2;1}, \pp_{3;1}, \pp_{1;12}, \pp_{2;12}, \pp_{3;12}  \}
\\
& \I_{2} : \{
\pp_{1}, \pp_{2}, \pp_{3}, \pp_{12}, \pp_{23}, \pp_{13}, \pp_{1;1}-m_V^2, \pp_{2;1}, \pp_{3;1}, \pp_{1;12}, \pp_{2;12}, \pp_{3;12}  \}
\\
& \I_{3} : \{
\pp_{1}-m_V^2, \pp_{2}, \pp_{3}, \pp_{12}, \pp_{23}, \pp_{13}, \pp_{1;1}, \pp_{2;1}, \pp_{3;1}, \pp_{1;12}, \pp_{2;12}, \pp_{3;12}  \}
\\
& \I_{4} : \{
\pp_{1}, \pp_{2}, \pp_{3}, \pp_{12}, \pp_{23}, \pp_{13}, \pp_{1;1}, \pp_{2;1}, \pp_{3;1}, \pp_{1;12}-m_V^2, \pp_{2;12}, \pp_{3;12}  \}
\\
& \I_{5} :  \{ 
\pp_{1}, \pp_{2}, \pp_{3}, \pp_{12}-m_V^2, \pp_{23}, \pp_{13}, \pp_{1;1}, \pp_{2;1}, \pp_{23;1}, \pp_{1;12}, \pp_{2;12}, \pp_{3;12}  \}
\\
& \I_{6} :  \{ 
\pp_{1}, \pp_{2}, \pp_{3}, \pp_{12}-m_V^2, \pp_{23}, \pp_{13}, \pp_{1;1}, \pp_{2;1}, \pp_{23;-2}, \pp_{1;12}, \pp_{2;12}, \pp_{3;12}  \}
\\
& \I_{7} :  \{ 
\pp_{1}, \pp_{2}, \pp_{3}, \pp_{12}, \pp_{23}, \pp_{13}, \pp_{1;1}, \pp_{2;1}, \pp_{23;1}-m_V^2, \pp_{1;12}, \pp_{2;12}, \pp_{3;12}  \}
\\
& \I_{8} :  \{ 
\pp_{1}, \pp_{2}, \pp_{3}, \pp_{12}, \pp_{23}, \pp_{13}, \pp_{1;1}, \pp_{2;1}, \pp_{23;-2}-m_V^2, \pp_{1;12}, \pp_{2;12}, \pp_{3;12}  \}
\\
& \I_{9} :  \{ 
\pp_{1}, \pp_{2}, \pp_{3}, \pp_{12}, \pp_{23}, \pp_{13}-m_V^2, \pp_{1;1}, \pp_{2;1}, \pp_{23;1}, \pp_{1;12}, \pp_{2;12}, \pp_{3;12}  \}
\\
& \I_{10} :  \{ 
\pp_{1}, \pp_{2}, \pp_{3}, \pp_{12}, \pp_{23}, \pp_{13}-m_V^2, \pp_{1;1}, \pp_{2;1}, \pp_{23;-2}, \pp_{1;12}, \pp_{2;12}, \pp_{3;12}  \}
\\
& \I_{11} :  \{ 
\pp_{1}, \pp_{2}, \pp_{3}, \pp_{12}, \pp_{23}, \pp_{13}, \pp_{1;1}, \pp_{2;1}-m_V^2, \pp_{23;1}, \pp_{1;12}, \pp_{2;12}, \pp_{3;12}  \}
\\
& \I_{12} :  \{ 
\pp_{1}, \pp_{2}, \pp_{3}, \pp_{12}, \pp_{23}, \pp_{13}, \pp_{1;1}, \pp_{2;1}-m_V^2, \pp_{23;-2}, \pp_{1;12}, \pp_{2;12}, \pp_{3;12}  \}
\\
& \I_{13} :  \{ 
\pp_{1}, \pp_{2}, \pp_{3}, \pp_{12}, \pp_{23}, \pp_{13}, \pp_{1;1}, \pp_{2;1}, \pp_{23;1}, \pp_{1;12}, \pp_{2;12}, \pp_{3;12}-m_V^2,
\\
& \I_{14} :  \{ 
\pp_{1}, \pp_{2}, \pp_{3}, \pp_{12}, \pp_{23}, \pp_{13}, \pp_{1;1}, \pp_{2;1}, \pp_{23;-2}, \pp_{1;12}, \pp_{2;12}, \pp_{3;12}-m_V^2,
\\
& \I_{15} :  \{ 
\pp_{1}, \pp_{2}, \pp_{3}-m_V^2, \pp_{12}, \pp_{23}, \pp_{13}, \pp_{1;1}, \pp_{2;1}, \pp_{23;1}, \pp_{1;12}, \pp_{2;12}, \pp_{3;12}  \}
\\
& \I_{16} :  \{ 
\pp_{1}, \pp_{2}, \pp_{3}, \pp_{12}, \pp_{23}, \pp_{13}, \pp_{1;1}-m_V^2, \pp_{2;1}, \pp_{23;1}, \pp_{1;12}, \pp_{2;12}, \pp_{3;12}  \}
\\
& \I_{17} :  \{ 
\pp_{1}, \pp_{2}, \pp_{3}, \pp_{12}, \pp_{23}, \pp_{13}, \pp_{1;1}-m_V^2, \pp_{2;1}, \pp_{23;-2}, \pp_{1;12}, \pp_{2;12}, \pp_{3;12}  \}
\\
& \I_{18} :  \{ 
\pp_{1}, \pp_{2}, \pp_{3}, \pp_{12}, \pp_{23}-m_V^2, \pp_{13}, \pp_{1;1}, \pp_{2;1}, \pp_{23;1}, \pp_{1;12}, \pp_{2;12}, \pp_{3;12}  \}
\\
& \I_{19} :  \{ 
\pp_{1}, \pp_{2}, \pp_{3}, \pp_{12}, \pp_{23}-m_V^2, \pp_{13}, \pp_{1;1}, \pp_{2;1}, \pp_{23;-2}, \pp_{1;12}, \pp_{2;12}, \pp_{3;12}  \}
\\
& \I_{20} :  \{ 
\pp_{1}, \pp_{2}-m_V^2, \pp_{3}, \pp_{12}, \pp_{23}, \pp_{13}, \pp_{1;1}, \pp_{2;1}, \pp_{23;-2}, \pp_{1;12}, \pp_{2;12}, \pp_{3;12}  \}
\\
& \I_{21} :  \{ 
\pp_{1}, \pp_{2}, \pp_{3}, \pp_{123}, \pp_{23}, \pp_{13}, \pp_{1;1}, \pp_{2;1}, \pp_{3;1}, \pp_{1;12}-m_V^2, \pp_{2;12}, \pp_{3;12}  \}
\\
& \I_{22} :  \{ 
\pp_{1}-m_V^2, \pp_{2}, \pp_{3}, \pp_{123}, \pp_{23}, \pp_{13}, \pp_{1;1}, \pp_{2;1}, \pp_{3;1}, \pp_{1;12}, \pp_{2;12}, \pp_{3;12}  \}
\\
& \I_{23} :  \{ 
\pp_{1}, \pp_{2}, \pp_{3}, \pp_{123}, \pp_{23}, \pp_{13}, \pp_{1;1}-m_V^2, \pp_{2;1}, \pp_{3;1}, \pp_{1;12}, \pp_{2;12}, \pp_{3;12}  \}
\\
& \I_{24} :  \{ 
\pp_{1}, \pp_{2}, \pp_{3}, \pp_{123}, \pp_{23}, \pp_{13}, \pp_{1;1}, \pp_{2;1}, \pp_{3;1}, \pp_{1;12}, \pp_{2;12}, \pp_{3;12}-m_V^2 \}
\\
& \I_{25} :  \{ 
\pp_{1}, \pp_{2}, \pp_{3}, \pp_{123}-m_V^2, \pp_{23}, \pp_{13}, \pp_{1;1}, \pp_{2;1}, \pp_{3;1}, \pp_{1;12}, \pp_{2;12}, \pp_{3;12}  \}
\end{align*}
We observe that certain integral families have sectors with high values. 
Therefore, to facilitate a smooth reduction using \textsc{Kira}, we have to optimize the ordering of the propagators.

\subsection{Master integrals}
Using \textsc{Kira}, we have performed the IBP reduction of the integral families mentioned above, which yielded a set of 303 MIs. 
We also have used \textsc{LiteRed}~\cite{Lee:2012cn,Lee:2013mka} to improve the reduction.
We now present the full list of MIs obtained in this work. 
To establish our notation, we first introduce the general form of a three-loop integral:
\begin{equation}
    \II_{n} (\nu_1,\nu_2,...,\nu_{12}) (d,x)= \int \prod_{i=1}^3 \frac{d^d k_i}{ (2 \pi)^{d}} \prod_{j=1}^{12} \frac{1}{D_{j}^{\nu_j}} \,,
\end{equation}
where $D_j$ is the $j^{th}$ element of the list given for $\II_{n}$.
We put $S_\ep = \exp ( - \ep (\gamma_E - \ln (4\pi)))$ equal to one for each loop order.
The MIs associated with the integral family $\II_{1}$ are as follows
\begin{align*}
\II_{1,1} &= \II_1(1,1,0,1,1,1,0,0,0,0,0,0). \qquad  &\II_{1,2} = \II_1(0,1,1,1,0,1,0,0,0,1,0,0). \\
\II_{1,3} &= \II_1(0,1,1,1,0,1,0,0,0,2,0,0). \qquad  &\II_{1,4} = \II_1(1,0,0,1,1,1,0,0,0,1,0,0). \\
\II_{1,5} &= \II_1(0,1,0,1,1,1,0,0,0,1,0,0). \qquad  &\II_{1,6} = \II_1(0,1,0,1,1,1,0,0,0,2,0,0). \\
\II_{1,7} &= \II_1(0,0,1,1,1,0,1,0,0,0,1,0). \qquad  &\II_{1,8} = \II_1(0,0,1,1,1,0,1,0,0,0,2,0). \\
\II_{1,9} &= \II_1(0,1,0,0,1,1,1,0,0,0,1,0). \qquad  &\II_{1,10} = \II_1(0,0,1,0,1,1,1,0,0,0,1,0). \\
\II_{1,11} &= \II_1(1,0,1,0,1,0,0,0,0,1,1,0). \qquad  &\II_{1,12} = \II_1(0,1,0,1,1,1,1,0,0,0,1,0). \\
\II_{1,13} &= \II_1(0,1,0,1,1,1,1,0,0,0,2,0). \qquad  &\II_{1,14} = \II_1(1,1,0,0,1,1,0,0,0,1,1,0). \\
\II_{1,15} &= \II_1(0,0,1,0,1,1,0,0,0,1,1,0). \qquad  &\II_{1,16} = \II_1(1,1,0,1,1,1,0,0,0,1,1,0). \\
\II_{1,17} &= \II_1(1,0,1,1,0,0,0,0,0,1,0,1). \qquad  &\II_{1,18} = \II_1(1,0,1,1,1,1,0,0,0,1,1,0). \\
\II_{1,19} &= \II_1(0,1,1,1,0,1,1,0,0,1,1,0). \qquad  &\II_{1,20} = \II_1(1,1,1,0,0,0,0,0,0,1,1,1). \\
\II_{1,21} &= \II_1(1,1,1,1,0,0,0,0,0,1,1,1). \qquad  &\II_{1,22} = \II_1(1,1,1,1,0,0,1,0,0,0,1,1). \\
\II_{1,23} &= \II_1(1,0,1,1,1,0,1,0,0,0,1,0). \qquad  &\II_{1,24} = \II_1(0,0,1,1,0,1,1,0,0,0,0,1). \\
\II_{1,25} &= \II_1(0,1,1,0,0,1,1,0,0,0,1,1). \qquad  &\II_{1,26} = \II_1(0,1,1,0,1,1,1,0,0,1,1,0). \\
\II_{1,27} &= \II_1(1,1,0,1,1,1,1,0,0,0,1,0). \qquad  &\II_{1,28} = \II_1(1,0,1,1,1,0,0,0,0,1,0,0). \\
\II_{1,29} &= \II_1(0,0,1,1,1,1,1,0,0,0,1,0). \qquad  &\II_{1,30} = \II_1(0,0,1,1,1,1,1,0,0,0,2,0). \\
\II_{1,31} &= \II_1(0,0,1,1,1,1,2,0,0,0,1,0). \qquad  &\II_{1,32} = \II_1(0,0,1,1,1,2,1,0,0,0,1,0). \\
\II_{1,33} &= \II_1(0,0,1,1,2,1,1,0,0,0,1,0). \qquad  &\II_{1,34} = \II_1(0,1,1,1,1,1,1,0,0,0,1,0). \\
\II_{1,35} &= \II_1(0,1,1,1,1,1,2,0,0,0,1,0). \qquad  &\II_{1,36} = \II_1(0,0,1,1,1,1,0,0,0,1,1,0). \\
\II_{1,37} &= \II_1(0,0,1,1,1,1,0,0,0,1,2,0). \qquad  &\II_{1,38} = \II_1(0,0,1,1,1,1,1,0,0,1,1,0). \\
\II_{1,39} &= \II_1(0,0,1,1,1,1,1,0,0,2,1,0). \qquad  &\II_{1,40} = \II_1(1,1,1,1,1,0,0,0,0,1,0,1). \\
\II_{1,41} &= \II_1(0,1,1,1,1,0,1,0,0,1,0,1). \qquad  &\II_{1,42} = \II_1(1,0,1,1,1,0,0,0,0,1,1,0). \\
\II_{1,43} &= \II_1(0,1,1,1,0,0,1,0,0,0,1,1). \qquad  &\II_{1,44} = \II_1(0,1,1,1,0,0,1,0,0,0,2,1). \\
\II_{1,45} &= \II_1(1,0,1,1,1,0,0,0,0,1,0,1). \qquad  &\II_{1,46} = \II_1(1,0,1,1,1,0,0,0,0,1,0,2). \\
\II_{1,47} &= \II_1(0,0,1,1,1,0,1,0,0,0,0,1). \qquad  &\II_{1,48} = \II_1(0,0,1,1,1,0,1,0,0,0,0,2). \\
\II_{1,49} &= \II_1(0,0,1,1,1,0,2,0,0,0,0,1). \qquad  &\II_{1,50} = \II_1(1,0,1,1,1,0,1,0,0,0,0,1). \\
\II_{1,51} &= \II_1(0,1,1,1,1,0,1,0,0,0,0,1). \qquad  &\II_{1,52} = \II_1(1,1,1,1,1,0,1,0,0,0,0,1). \\
\II_{1,53} &= \II_1(0,0,1,1,0,1,0,0,0,1,0,0). \qquad  &\II_{1,54} = \II_1(0,0,1,1,1,0,0,0,0,1,0,0). \\
\II_{1,55} &= \II_1(0,0,1,1,1,0,0,0,0,2,0,0). \qquad  &\II_{1,56} = \II_1(0,1,1,1,0,1,1,0,0,0,1,0). \\
\II_{1,57} &= \II_1(0,1,1,1,0,1,1,0,0,0,2,0). \qquad  &\II_{1,58} = \II_1(0,1,0,0,1,1,0,0,0,1,0,0). \\
\II_{1,59} &= \II_1(0,1,1,1,0,1,1,0,0,0,0,1). \qquad  &\II_{1,60} = \II_1(0,1,1,1,0,1,1,0,0,0,0,2). \\
\II_{1,61} &= \II_1(0,1,1,1,0,1,2,0,0,0,0,1). \qquad  &\II_{1,62} = \II_1(1,0,1,1,1,0,0,0,0,0,0,0). \\
\II_{1,63} &= \II_1(0,1,1,1,0,0,0,0,0,1,0,1). \qquad  &\II_{1,64} = \II_1(0,1,1,1,0,0,0,0,0,2,0,1). \\
\II_{1,65} &= \II_1(0,1,1,1,0,1,0,0,1,1,0,0). \qquad  &\II_{1,66} = \II_1(0,1,1,1,0,1,0,0,1,1,1,0). \\
\II_{1,67} &= \II_1(0,1,1,1,0,1,0,0,1,1,2,0). \qquad  &\II_{1,68} = \II_1(1,1,0,1,0,1,0,0,1,0,1,0). \\
\II_{1,69} &= \II_1(1,1,0,1,0,1,0,0,1,1,1,0). \qquad  &\II_{1,70} = \II_1(0,1,0,1,0,1,0,0,1,1,0,0). \\
\II_{1,71} &= \II_1(0,1,0,1,0,1,0,0,1,2,0,0). \qquad  &\II_{1,72} = \II_1(1,1,0,1,1,1,0,0,1,1,0,0). \\
\II_{1,73} &= \II_1(1,1,0,1,1,1,0,0,1,2,0,0). \qquad  &\II_{1,74} = \II_1(1,1,0,1,1,1,0,0,2,1,0,0). \\
\II_{1,75} &= \II_1(1,1,0,0,1,1,0,0,1,1,0,0). \qquad  &\II_{1,76} = \II_1(0,1,0,1,1,1,0,0,1,1,0,0). \\
\II_{1,77} &= \II_1(0,1,0,1,1,1,0,0,1,2,0,0). \qquad  &\II_{1,78} = \II_1(0,1,0,1,1,1,0,0,2,1,0,0). \\
\II_{1,79} &= \II_1(0,1,0,1,1,2,0,0,1,1,0,0). \qquad  &\II_{1,80} = \II_1(0,1,0,1,1,0,1,0,1,0,1,0). \\
\II_{1,81} &= \II_1(0,1,0,1,1,0,1,0,1,0,2,0). \qquad  &\II_{1,82} = \II_1(0,0,1,1,1,0,1,0,1,0,1,0). \\
\II_{1,83} &= \II_1(0,1,1,1,1,1,1,0,1,1,1,0). \qquad  &\II_{1,84} = \II_1(0,1,1,1,1,1,1,0,1,1,2,0). \\
\II_{1,85} &= \II_1(0,1,1,1,1,1,1,0,2,1,1,0). \qquad  &\II_{1,86} = \II_1(1,0,1,1,1,0,1,1,0,0,0,1). \\
\II_{1,87} &= \II_1(0,0,1,0,1,1,1,1,0,0,0,1). \qquad  &\II_{1,88} = \II_1(0,0,1,1,1,1,1,1,0,0,0,1). \\
\II_{1,89} &= \II_1(0,0,1,1,1,1,1,1,0,0,0,2). \qquad  &\II_{1,90} = \II_1(0,0,1,1,1,1,1,1,0,1,0,0). \\
\II_{1,91} &= \II_1(1,0,1,1,1,1,0,1,0,0,0,1). \qquad  &\II_{1,92} = \II_1(0,1,0,1,1,1,1,1,0,1,0,0). \\
\II_{1,93} &= \II_1(0,1,1,1,0,0,1,1,0,1,0,1). \qquad  &\II_{1,94} = \II_1(0,1,1,1,1,1,1,1,0,1,0,1). \\
\II_{1,95} &= \II_1(0,1,1,1,1,1,1,1,0,1,0,2). \qquad  &\II_{1,96} = \II_1(0,1,1,1,1,1,0,0,0,1,0,1). 
%
%
\end{align*}    
The MIs belonging to the integral family $\II_{2}$ are: 
\begin{align*}
 \II_{2,1} &= \II_2(1,0,0,0,1,1,1,1,0,1,0,0). \qquad  &\II_{2,2} = \II_2(1,0,0,0,1,1,1,1,0,0,1,0). \\
 \II_{2,3} &= \II_2(1,0,0,1,0,1,1,1,0,0,0,1). \qquad  &\II_{2,4} = \II_2(1,0,0,0,1,1,1,1,0,0,0,1). \\
 \II_{2,5} &= \II_2(1,0,0,0,1,1,1,1,0,0,0,2). \qquad  &\II_{2,6} = \II_2(1,0,0,1,0,1,1,0,0,0,1,1). \\
 \II_{2,7} &= \II_2(1,0,0,1,0,1,1,1,0,0,1,1). \qquad  &\II_{2,8} = \II_2(1,0,0,1,1,0,1,1,0,1,0,1). \\
 \II_{2,9} &= \II_2(1,0,1,1,1,0,1,0,0,0,1,0). \qquad  &\II_{2,10} = \II_2(1,0,2,1,1,0,1,0,0,0,1,0). \\
 \II_{2,11} &= \II_2(1,0,1,1,1,0,1,0,0,1,0,1). \qquad  &\II_{2,12} = \II_2(1,0,1,1,0,0,1,1,0,1,0,1). \\
 \II_{2,13} &= \II_2(1,0,1,1,0,0,1,0,0,0,1,1). \qquad  &\II_{2,14} = \II_2(0,0,1,1,0,1,1,0,0,0,1,1). \\
 \II_{2,15} &= \II_2(0,0,2,1,0,1,1,0,0,0,1,1). \qquad  &\II_{2,16} = \II_2(1,0,1,1,1,0,1,0,0,0,1,1). \\
 \II_{2,17} &= \II_2(1,0,1,1,1,0,1,0,0,0,1,2). \qquad  &\II_{2,18} = \II_2(1,0,1,1,0,1,1,0,0,0,1,1). \\
 \II_{2,19} &= \II_2(1,0,1,1,0,0,1,1,0,0,1,1). \qquad  &\II_{2,20} = \II_2(1,0,1,0,1,0,1,0,0,1,1,0). \\
 \II_{2,21} &= \II_2(0,0,1,1,0,1,1,1,0,0,1,1). \qquad  &\II_{2,22} = \II_2(0,0,2,1,0,1,1,1,0,0,1,1). \\
 \II_{2,23} &= \II_2(1,0,1,1,1,0,1,1,0,0,0,1). \qquad  &\II_{2,24} = \II_2(1,0,2,1,1,0,1,1,0,0,0,1). \\
 \II_{2,25} &= \II_2(1,0,1,1,1,0,1,1,0,0,0,2). \qquad  &\II_{2,26} = \II_2(1,0,1,1,0,1,1,1,0,0,0,1). \\
 \II_{2,27} &= \II_2(1,0,1,1,1,1,1,1,0,0,1,1). \qquad  &\II_{2,28} = \II_2(1,0,2,1,1,1,1,1,0,0,1,1). \\
 \II_{2,29} &= \II_2(1,0,1,1,1,1,1,1,0,0,1,2). \qquad  &\II_{2,30} = \II_2(0,0,1,1,0,1,1,0,0,0,1,0). \\
 \II_{2,31} &= \II_2(0,0,2,1,0,1,1,0,0,0,1,0). \qquad  &\II_{2,32} = \II_2(0,0,1,1,1,1,1,0,0,0,1,0). \\
 \II_{2,33} &= \II_2(0,0,2,1,1,1,1,0,0,0,1,0). \qquad  &\II_{2,34} = \II_2(0,0,1,1,1,1,2,0,0,0,1,0). \\
 \II_{2,35} &= \II_2(1,0,1,1,1,0,1,0,0,0,0,1). \qquad  &\II_{2,36} = \II_2(1,0,0,1,1,0,1,1,0,0,0,1). \\
 \II_{2,37} &= \II_2(1,0,1,0,1,1,1,0,0,0,1,0). \qquad  &\II_{2,38} = \II_2(1,0,0,0,1,1,1,0,0,0,1,0). \\
 \II_{2,39} &= \II_2(1,0,0,1,1,1,0,1,0,0,0,1). \qquad  &\II_{2,40} = \II_2(1,0,0,1,1,1,1,1,0,0,0,1). \\
 \II_{2,41} &= \II_2(1,0,0,1,1,1,1,1,0,0,0,2). \qquad  &\II_{2,42} = \II_2(1,1,1,0,0,0,1,0,0,1,1,1). \\
 \II_{2,43} &= \II_2(1,0,0,1,0,1,1,1,1,1,0,0). \qquad  &\II_{2,44} = \II_2(0,0,1,1,0,1,1,1,1,0,1,0). \\
 \II_{2,45} &= \II_2(0,0,1,1,0,1,1,1,1,1,0,0). \qquad  &
\end{align*}
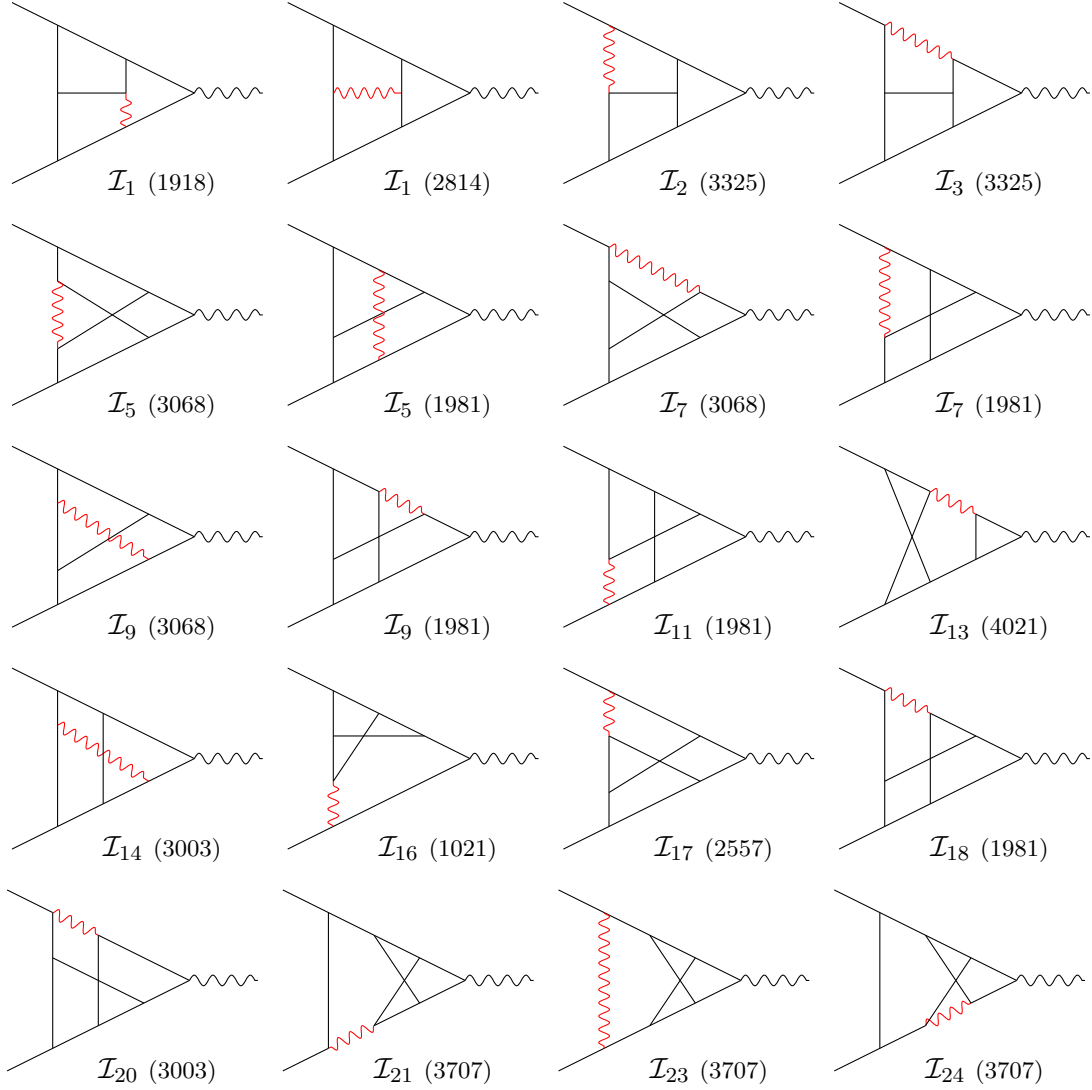
\begin{figure}[h] \centering
%
\begin{tikzpicture}[scale=0.6]
 \draw[scalar]  (-4,2) -- (-3,1.5);
 \draw[scalar]  (-1.5,0.75) -- (0,0);
 \draw[scalar]  (0,0) -- (-4,-2);
 \draw[photon] (0,0) -- (1.5,0); 
 \draw[scalar] (-3,1.5) -- (-1.5,0.75);
 \draw[scalar] (-3,1.5) -- (-3,-1.5);
 \draw[scalar] (-3,0) -- (-1.5,0);
 \draw[scalar] (-1.5,0.75) -- (-1.5,0);
 \draw[Zboson] (-1.5,-0.75) -- (-1.5,0);
 \node at (1.5,0) {$~$};
 \node at (-0.7,-2) {$\I_1 ~ \text{\footnotesize (1918)}$};
\end{tikzpicture}
%
\begin{tikzpicture}[scale=0.6]
 \draw[scalar]  (-4,2) -- (-3,1.5);
 \draw[scalar]  (-1.5,0.75) -- (0,0);
 \draw[scalar]  (0,0) -- (-4,-2);
 \draw[photon] (0,0) -- (1.5,0); 
 \draw[scalar] (-3,1.5) -- (-1.5,0.75);
 \draw[scalar] (-3,1.5) -- (-3,-1.5);
 \draw[Zboson] (-3,0) -- (-1.5,0);
 \draw[scalar] (-1.5,0.75) -- (-1.5,-0.75);
 \node at (1.5,0) {$~$};
 \node at (-0.7,-2) {$\I_1 ~ \text{\footnotesize (2814)}$};
\end{tikzpicture}
%
\begin{tikzpicture}[scale=0.6]
 \draw[scalar]  (-4,2) -- (-3,1.5);
 \draw[scalar]  (-1.5,0.75) -- (0,0);
 \draw[scalar]  (0,0) -- (-4,-2);
 \draw[photon] (0,0) -- (1.5,0); 
 \draw[scalar] (-3,1.5) -- (-1.5,0.75);
 \draw[scalar] (-3,0) -- (-3,-1.5);
 \draw[Zboson] (-3,1.5) -- (-3,0);
 \draw[scalar] (-3,0) -- (-1.5,0);
 \draw[scalar] (-1.5,0.75) -- (-1.5,-0.75);
 \node at (1.5,0) {$~$};
 \node at (-0.7,-2) {$\I_2 ~ \text{\footnotesize (3325)}$};
\end{tikzpicture}
%
\begin{tikzpicture}[scale=0.6]
 \draw[scalar]  (-4,2) -- (-3,1.5);
 \draw[scalar]  (-1.5,0.75) -- (0,0);
 \draw[scalar]  (0,0) -- (-4,-2);
 \draw[photon] (0,0) -- (1.5,0); 
 \draw[Zboson] (-3,1.5) -- (-1.5,0.75);
 \draw[scalar] (-3,1.5) -- (-3,-1.5);
 \draw[scalar] (-3,0) -- (-1.5,0);
 \draw[scalar] (-1.5,0.75) -- (-1.5,-0.75);
 \node at (1.5,0) {$~$};
 \node at (-0.7,-2) {$\I_3 ~ \text{\footnotesize (3325)}$};
\end{tikzpicture}

\vspace{0.2cm}

\begin{tikzpicture}[scale=0.6]
 \draw[scalar]  (-4,2) -- (0,0);
 \draw[scalar]  (0,0) -- (-4,-2);
 \draw[photon]  (0,0) -- (1.5,0); 
 \draw[scalar]  (-3,1.5) -- (-3,0.75);
 \draw[scalar] (-3,0.75) -- (-1,-0.5);
 \draw[scalar] (-3,-0.75) -- (-1,0.5);
 \draw[scalar] (-3,-0.75) -- (-3,-1.5);
 \draw[Zboson] (-3,0.75) -- (-3,-0.75);
 \node at (1.5,0) {$~$};
 \node at (-0.7,-2) {$\I_5 ~ \text{\footnotesize (3068)}$};
\end{tikzpicture}
%
\begin{tikzpicture}[scale=0.6]
 \draw[scalar]  (-4,2) -- (0,0);
 \draw[scalar]  (0,0) -- (-4,-2);
 \draw[photon] (0,0) -- (1.5,0); 
 \draw[scalar] (-3,1.5) -- (-3,-1.5);
 \draw[scalar] (-3,-0.5) -- (-1,0.5);
 \draw[Zboson] (-2,1) -- (-2,-1);
 \node at (1.5,0) {$~$};
 \node at (-0.7,-2) {$\I_5 ~ \text{\footnotesize (1981)}$};
\end{tikzpicture}
%
\begin{tikzpicture}[scale=0.6]
 \draw[scalar]  (-4,2) -- (-3,1.5);
 \draw[scalar]  (0,0) -- (-4,-2);
 \draw[photon]  (0,0) -- (1.5,0); 
 \draw[scalar]  (-3,1.5) -- (-3,-1.5);
 \draw[scalar] (-3,0.75) -- (-1,-0.5);
 \draw[scalar] (-3,-0.75) -- (-1,0.5);
 \draw[scalar] (-1,0.5) -- (0,0);
 \draw[Zboson] (-3,1.5) -- (-1,0.5);
 \node at (1.5,0) {$~$};
 \node at (-0.7,-2) {$\I_7 ~ \text{\footnotesize (3068)}$};
\end{tikzpicture}
%
\begin{tikzpicture}[scale=0.6]
 \draw[scalar]  (-4,2) -- (0,0);
 \draw[scalar]  (0,0) -- (-4,-2);
 \draw[photon] (0,0) -- (1.5,0); 
 \draw[Zboson] (-3,1.5) -- (-3,-0.5);
 \draw[scalar] (-3,-0.5) -- (-3,-1.5);
 \draw[scalar] (-3,-0.5) -- (-1,0.5);
 \draw[scalar] (-2,1) -- (-2,-1);
 \node at (1.5,0) {$~$};
 \node at (-0.7,-2) {$\I_7 ~ \text{\footnotesize (1981)}$};
\end{tikzpicture}

\vspace{0.2cm}

\begin{tikzpicture}[scale=0.6]
 \draw[scalar]  (-4,2) -- (0,0);
 \draw[scalar]  (0,0) -- (-4,-2);
 \draw[photon]  (0,0) -- (1.5,0); 
 \draw[scalar]  (-3,1.5) -- (-3,-1.5);
 \draw[Zboson] (-3,0.75) -- (-1,-0.5);
 \draw[scalar] (-3,-0.75) -- (-1,0.5);
 \node at (1.5,0) {$~$};
 \node at (-0.7,-2) {$\I_9 ~ \text{\footnotesize (3068)}$};
\end{tikzpicture}
%
\begin{tikzpicture}[scale=0.6]
 \draw[scalar]  (-4,2) -- (-2,1);
 \draw[scalar]  (-1,0.5) -- (0,0);
 \draw[scalar]  (0,0) -- (-4,-2);
 \draw[scalar]  (-2,1) -- (-2,-1);
 \draw[photon]  (0,0) -- (1.5,0); 
 \draw[scalar]  (-3,1.5) -- (-3,-1.5);
 \draw[scalar] (-3,-0.5) -- (-1,0.5);
 \draw[Zboson] (-2,1) -- (-1,0.5);
 \node at (1.5,0) {$~$};
 \node at (-0.7,-2) {$\I_9 ~ \text{\footnotesize (1981)}$};
\end{tikzpicture}
%
\begin{tikzpicture}[scale=0.6]
 \draw[scalar]  (-4,2) -- (0,0);
 \draw[scalar]  (0,0) -- (-4,-2);
 \draw[scalar]  (-2,1) -- (-2,-1);
 \draw[photon]  (0,0) -- (1.5,0); 
 \draw[scalar]  (-3,1.5) -- (-3,-0.5);
 \draw[scalar] (-3,-0.5) -- (-1,0.5);
 \draw[Zboson] (-3,-1.5) -- (-3,-0.5);
 \node at (1.5,0) {$~$};
 \node at (-0.7,-2) {$\I_{11} ~ \text{\footnotesize (1981)}$};
\end{tikzpicture}
%
\begin{tikzpicture}[scale=0.6]
 \draw[scalar]  (-4,2) -- (-2,1);
 \draw[scalar]  (-1,0.5) -- (0,0);
 \draw[scalar]  (0,0) -- (-4,-2);
 \draw[scalar]  (-2,1) -- (-3,-1.5);
 \draw[photon]  (0,0) -- (1.5,0); 
 \draw[scalar]  (-3,1.5) -- (-2,-1);
 \draw[scalar] (-1,-0.5) -- (-1,0.5);
 \draw[Zboson] (-2,1) -- (-1,0.5);
 \node at (1.5,0) {$~$};
 \node at (-0.7,-2) {$\I_{13} ~ \text{\footnotesize (4021)}$};
\end{tikzpicture}

\vspace{0.2cm}

\begin{tikzpicture}[scale=0.6]
 \draw[scalar]  (-4,2) -- (0,0);
 \draw[scalar]  (0,0) -- (-4,-2);
 \draw[photon]  (0,0) -- (1.5,0); 
 \draw[scalar]  (-3,1.5) -- (-3,-1.5);
 \draw[Zboson] (-3,0.75) -- (-1,-0.5);
 \draw[scalar] (-2,-1) -- (-2,1);
 \node at (1.5,0) {$~$};
 \node at (-0.7,-2) {$\I_{14} ~ \text{\footnotesize (3003)}$};
\end{tikzpicture}
%
\begin{tikzpicture}[scale=0.6]
 \draw[scalar]  (-4,2) -- (0,0);
 \draw[scalar]  (0,0) -- (-4,-2);
 \draw[scalar]  (-3,0.5) -- (-1,0.5);
 \draw[photon]  (0,0) -- (1.5,0); 
 \draw[scalar]  (-3,1.5) -- (-3,-0.5);
 \draw[scalar] (-3,-0.5) -- (-2,1);
 \draw[Zboson] (-3,-1.5) -- (-3,-0.5);
 \node at (1.5,0) {$~$};
 \node at (-0.7,-2) {$\I_{16} ~ \text{\footnotesize (1021)}$};
\end{tikzpicture}
%
\begin{tikzpicture}[scale=0.6]
 \draw[scalar]  (-4,2) -- (0,0);
 \draw[scalar]  (0,0) -- (-4,-2);
 \draw[photon]  (0,0) -- (1.5,0); 
 \draw[scalar]  (-3,0.5) -- (-3,-1.5);
 \draw[scalar] (-3,0.5) -- (-1,-0.5);
 \draw[scalar] (-3,-0.75) -- (-1,0.5);
 \draw[Zboson] (-3,1.5) -- (-3,0.5);
 \node at (1.5,0) {$~$};
 \node at (-0.7,-2) {$\I_{17} ~ \text{\footnotesize (2557)}$};
\end{tikzpicture}
%
\begin{tikzpicture}[scale=0.6]
 \draw[scalar]  (-4,2) -- (-3,1.5);
 \draw[scalar]  (-2,1) -- (0,0);
 \draw[scalar]  (0,0) -- (-4,-2);
 \draw[scalar]  (-2,1) -- (-2,-1);
 \draw[photon]  (0,0) -- (1.5,0); 
 \draw[scalar]  (-3,1.5) -- (-3,-1.5);
 \draw[scalar] (-3,-0.5) -- (-1,0.5);
 \draw[Zboson] (-3,1.5) -- (-2,1);
 \node at (1.5,0) {$~$};
 \node at (-0.7,-2) {$\I_{18} ~ \text{\footnotesize (1981)}$};
\end{tikzpicture}

\vspace{0.2cm}

\begin{tikzpicture}[scale=0.6]
 \draw[scalar]  (-4,2) -- (-3,1.5);
 \draw[scalar]  (-2,1) -- (0,0);
 \draw[scalar]  (0,0) -- (-4,-2);
 \draw[scalar]  (-2,1) -- (-2,-1);
 \draw[photon]  (0,0) -- (1.5,0); 
 \draw[scalar]  (-3,1.5) -- (-3,-1.5);
 \draw[scalar] (-3,0.5) -- (-1,-0.5);
 \draw[Zboson] (-3,1.5) -- (-2,1);
 \node at (1.5,0) {$~$};
 \node at (-0.7,-2) {$\I_{20} ~ \text{\footnotesize (3003)}$};
\end{tikzpicture}
%
\begin{tikzpicture}[scale=0.6]
 \draw[scalar]  (-4,2) -- (0,0);
 \draw[scalar]  (0,0) -- (-2,-1);
 \draw[scalar]  (-4,-2) -- (-3,-1.5);
 \draw[Zboson]  (-2,-1) -- (-3,-1.5); 
 \draw[photon] (0,0) -- (1.5,0); 
 \draw[scalar] (-3,1.5) -- (-3,-1.5);
 \draw[scalar] (-2,1.0) -- (-1,-0.5);
 \draw[scalar] (-1,0.5) -- (-2,-1.0);
 \node at (1.5,0) {$~$};
 \node at (-0.7,-2) {$\I_{21} ~ \text{\footnotesize (3707)}$};
\end{tikzpicture}
%
\begin{tikzpicture}[scale=0.6]
 \draw[scalar]  (-4,2) -- (0,0);
 \draw[scalar]  (0,0) -- (-4,-2);
 \draw[photon] (0,0) -- (1.5,0); 
 \draw[Zboson] (-3,1.5) -- (-3,-1.5);
 \draw[scalar] (-2,1.0) -- (-1,-0.5);
 \draw[scalar] (-1,0.5) -- (-2,-1.0);
 \node at (1.5,0) {$~$};
 \node at (-0.7,-2) {$\I_{23} ~ \text{\footnotesize (3707)}$};
\end{tikzpicture}
%
\begin{tikzpicture}[scale=0.6]
 \draw[scalar]  (-4,2) -- (0,0);
 \draw[scalar]  (0,0) -- (-1,-0.5);
 \draw[scalar]  (-4,-2) -- (-2,-1);
 \draw[Zboson]  (-2,-1) -- (-1,-0.5); 
 \draw[photon] (0,0) -- (1.5,0); 
 \draw[scalar] (-3,1.5) -- (-3,-1.5);
 \draw[scalar] (-2,1.0) -- (-1,-0.5);
 \draw[scalar] (-1,0.5) -- (-2,-1.0);
 \node at (1.5,0) {$~$};
 \node at (-0.7,-2) {$\I_{24} ~ \text{\footnotesize (3707)}$};
\end{tikzpicture}
\caption{Schematic representation of the appearing nine-propagator master integrals for each integral family. The red wavy line represents the massive propagator.}
\label{fig:mis}
\end{figure}

The integral family $\II_{3}$ has the following 51 MIs
\begin{align*}
\II_{3,1} &= \II_3(1,1,0,1,1,0,1,0,1,0,1,0). \qquad  &\II_{3,2} = \II_3(1,1,0,1,1,0,1,0,1,0,2,0). \\
\II_{3,3} &= \II_3(1,0,0,1,0,1,0,0,1,0,1,0). \qquad  &\II_{3,4} = \II_3(1,0,0,1,0,1,0,0,1,0,2,0). \\
\II_{3,5} &= \II_3(1,1,0,1,0,1,0,0,1,0,1,0). \qquad  &\II_{3,6} = \II_3(1,1,0,1,0,1,0,0,1,0,2,0). \\
\II_{3,7} &= \II_3(1,1,1,1,0,1,0,0,1,0,1,0). \qquad  &\II_{3,8} = \II_3(1,1,1,1,0,1,0,0,1,0,2,0). \\
\II_{3,9} &= \II_3(1,1,1,0,1,1,1,0,0,0,1,0). \qquad  &\II_{3,10} = \II_3(1,1,1,0,1,1,2,0,0,0,1,0). \\
\II_{3,11} &= \II_3(1,1,0,1,1,1,0,0,1,0,1,0). \qquad  &\II_{3,12} = \II_3(1,1,0,0,1,1,1,0,0,0,1,0). \\
\II_{3,13} &= \II_3(1,1,0,0,1,1,1,0,0,0,2,0). \qquad  &\II_{3,14} = \II_3(1,0,1,0,1,1,1,0,0,0,1,0). \\
\II_{3,15} &= \II_3(1,0,1,0,1,1,1,0,0,0,2,0). \qquad  &\II_{3,16} = \II_3(1,0,1,1,1,1,1,0,0,0,1,0). \\
\II_{3,17} &= \II_3(1,0,1,1,1,1,1,0,0,0,2,0). \qquad  &\II_{3,18} = \II_3(1,0,0,1,1,1,0,0,1,0,1,0). \\
\II_{3,19} &= \II_3(1,0,0,1,1,1,0,0,1,0,2,0). \qquad  &\II_{3,20} = \II_3(1,0,1,1,1,0,1,0,0,0,1,0). \\
\II_{3,21} &= \II_3(1,0,1,1,1,0,1,0,0,0,2,0). \qquad  &\II_{3,22} = \II_3(1,1,1,1,0,1,1,0,0,0,1,0). \\
\II_{3,23} &= \II_3(1,1,1,1,0,1,1,0,0,0,2,0). \qquad  &\II_{3,24} = \II_3(1,0,0,1,0,1,0,0,0,0,1,1). \\
\II_{3,25} &= \II_3(1,0,0,1,0,1,0,0,0,0,1,2). \qquad  &\II_{3,26} = \II_3(1,1,0,0,1,1,1,0,0,0,1,1). \\
\II_{3,27} &= \II_3(1,1,0,0,1,1,1,0,0,0,1,2). \qquad  &\II_{3,28} = \II_3(1,1,0,1,0,1,0,0,1,0,1,1). \\
\II_{3,29} &= \II_3(1,1,0,1,0,1,0,0,1,0,1,2). \qquad  &\II_{3,30} = \II_3(1,1,0,1,0,1,1,0,0,0,1,1). \\
\II_{3,31} &= \II_3(1,1,0,1,0,1,1,0,0,0,1,2). \qquad  &\II_{3,32} = \II_3(1,1,0,0,1,1,1,0,1,0,1,0). \\
\II_{3,33} &= \II_3(1,1,0,1,1,1,1,0,1,0,1,1). \qquad  &\II_{3,34} = \II_3(1,1,0,1,1,1,1,0,1,0,1,2). \\
\II_{3,35} &= \II_3(1,1,0,1,1,0,1,0,1,0,0,1). \qquad  &\II_{3,36} = \II_3(1,1,1,1,0,1,0,0,0,0,1,1). \\
\II_{3,37} &= \II_3(1,1,1,1,0,1,0,0,0,0,1,2). \qquad  &\II_{3,38} = \II_3(1,1,1,1,0,0,1,0,0,0,1,1). \\
\II_{3,39} &= \II_3(1,1,1,1,0,0,1,0,0,0,2,1). \qquad  &\II_{3,40} = \II_3(1,1,1,1,0,1,0,0,0,1,0,0). \\
\II_{3,41} &= \II_3(1,0,1,1,1,0,0,0,0,1,0,0). \qquad  &\II_{3,42} = \II_3(1,1,1,0,0,1,0,0,0,1,1,0). \\
\II_{3,43} &= \II_3(1,1,1,0,0,0,0,0,0,1,1,1). \qquad  &\II_{3,44} = \II_3(1,0,1,0,1,0,0,0,0,1,1,0). \\
\II_{3,45} &= \II_3(1,1,0,0,1,1,0,0,0,1,1,0). \qquad  &\II_{3,46} = \II_3(1,0,1,0,1,1,0,0,0,1,1,0). \\
\II_{3,47} &= \II_3(1,1,1,0,1,1,0,0,0,1,1,0). \qquad  &\II_{3,48} = \II_3(1,0,1,1,0,1,0,1,0,1,0,0). \\
\II_{3,49} &= \II_3(1,1,1,1,0,1,0,1,0,1,0,0). \qquad  &\II_{3,50} = \II_3(1,0,1,1,0,0,0,1,0,1,0,1). \\
\II_{3,51} &= \II_3(1,1,1,1,0,0,0,1,0,1,0,1). \qquad  &
\end{align*}
The MIs belonging to the integral families $\II_{5}$, $\II_{7}$ and $\II_{9}$ are
\begin{align*}
\II_{5,1} &= \II_5(0,0,1,1,1,0,1,0,1,1,0,0). \qquad  &\II_{5,2} = \II_5(0,0,1,1,1,0,1,0,1,0,1,1). \\
\II_{5,3} &= \II_5(0,0,1,1,1,0,1,0,1,0,2,1). \qquad  &\II_{5,4} = \II_5(0,0,1,1,1,0,0,1,1,0,1,1). \\
\II_{5,5} &= \II_5(0,0,1,1,1,0,1,0,1,1,0,1). \qquad  &\II_{5,6} = \II_5(0,0,1,1,1,0,1,0,1,2,0,1). \\
\II_{5,7} &= \II_5(0,0,1,1,0,1,0,1,1,1,0,1). \qquad  &\II_{5,8} = \II_5(0,0,1,1,0,1,0,1,1,2,0,1). \\
\II_{5,9} &= \II_5(0,0,1,1,1,1,1,1,1,1,0,1). \qquad  &\II_{5,10} = \II_5(0,0,1,1,1,1,1,1,1,2,0,1). \\
\II_{5,11} &= \II_5(0,0,1,1,1,1,1,1,1,1,0,2). \qquad  &\II_{5,12} = \II_5(1,0,0,1,1,1,0,1,1,0,1,0). \\
\II_{5,13} &= \II_5(0,0,1,0,1,1,0,1,1,1,1,0). \qquad  &\II_{5,14} = \II_5(1,0,1,1,1,1,0,1,1,1,0,0). \\
\II_{5,15} &= \II_5(1,0,1,1,1,1,0,1,1,2,0,0). \qquad  &\II_{5,16} = \II_5(1,0,1,1,0,1,0,0,1,1,1,0). \\
\II_{5,17} &= \II_5(1,0,1,1,0,1,0,0,1,1,2,0). \qquad  &\II_{5,18} = \II_5(1,0,0,1,0,1,0,1,1,1,1,0). \\
\II_{5,19} &= \II_5(1,0,0,1,0,1,0,1,1,1,2,0). \qquad  &\II_{5,20} = \II_5(1,0,1,1,1,1,0,0,1,0,1,0). \\
\II_{5,21} &= \II_5(1,0,0,0,1,1,0,1,1,1,1,0). \qquad  &\II_{5,22} = \II_5(0,0,1,1,1,1,0,1,1,1,1,0). \\
\II_{5,23} &= \II_5(0,0,1,1,1,1,0,1,1,1,2,0). \qquad  &\II_{5,24} = \II_5(1,0,1,1,1,1,0,0,1,1,1,0). \\
\II_{5,25} &= \II_5(1,0,0,1,1,1,0,1,1,1,1,0). \qquad  &\II_{5,26} = \II_5(1,0,1,1,1,1,0,1,1,1,1,0). \\
\II_{5,27} &= \II_5(1,0,1,1,1,1,0,1,1,1,2,0). \qquad  &\II_{5,28} = \II_5(1,0,1,1,1,1,0,1,1,2,1,0). \\
%
%
\II_{7,1} &= \II_7(1,0,0,0,1,1,0,1,1,1,1,0). \qquad  &\II_{7,2} = \II_7(1,0,0,0,1,1,0,1,1,1,2,0). \\
\II_{7,3} &= \II_7(0,0,1,0,1,1,0,1,1,1,1,0). \qquad  &\II_{7,4} = \II_7(0,0,1,0,1,1,0,1,1,1,2,0). \\
\II_{7,5} &= \II_7(0,0,1,1,1,1,0,1,1,1,0,0). \qquad  &\II_{7,6} = \II_7(0,0,1,1,1,1,0,1,1,2,0,0). \\
\II_{7,7} &= \II_7(1,0,1,1,1,1,0,1,1,1,0,0). \qquad  &\II_{7,8} = \II_7(1,0,1,1,1,1,0,1,1,2,0,0). \\
\II_{7,9} &= \II_7(1,0,1,1,1,1,0,0,1,1,1,0). \qquad  &\II_{7,10} = \II_7(1,0,1,0,1,1,0,1,1,1,1,0). \\
\II_{7,11} &= \II_7(1,0,1,1,1,1,0,1,1,1,1,0). \qquad  &\II_{7,12} = \II_7(1,0,1,1,1,1,0,1,1,1,2,0). \\
\II_{7,13} &= \II_7(1,0,1,1,1,1,0,1,1,2,1,0). \qquad  &\II_{7,14} = \II_7(0,0,1,1,1,0,0,1,1,1,0,1). \\
\II_{7,15} &= \II_7(0,0,1,1,1,0,0,1,1,1,0,2). \qquad  &\II_{7,16} = \II_7(0,0,1,1,1,0,0,1,1,2,0,1). \\
\II_{7,17} &= \II_7(0,0,0,1,0,1,1,1,1,1,0,1). \qquad  &\II_{7,18} = \II_7(0,0,1,1,1,0,1,0,1,0,1,1). \\
\II_{7,19} &= \II_7(0,0,1,1,1,0,1,0,1,0,2,1). \qquad  &\II_{7,20} = \II_7(0,0,1,1,1,1,1,1,1,1,0,1). \\
\II_{7,21} &= \II_7(0,0,1,1,1,2,1,1,1,1,0,1). \qquad  &\II_{7,22} = \II_7(0,0,1,1,1,1,1,1,1,1,0,2). \\
\II_{7,23} &= \II_7(1,0,0,0,1,1,0,1,1,0,1,1). \qquad  &\II_{7,24} = \II_7(1,0,1,0,1,0,0,0,1,1,1,1). \\
\II_{7,25} &= \II_7(1,0,1,0,1,0,0,1,1,1,1,1). \qquad  &\II_{7,26} = \II_7(1,0,0,0,1,1,0,1,1,1,1,1). \\
%
%
\II_{9,1} &= \II_9(0,0,1,1,1,1,0,0,1,1,1,0). \qquad  &\II_{9,2} = \II_9(1,0,1,1,1,1,0,1,1,1,0,0). \\
\II_{9,3} &= \II_9(1,0,1,1,1,1,0,1,1,2,0,0). \qquad  &\II_{9,4} = \II_9(0,0,1,0,1,1,0,1,1,1,1,0). \\
\II_{9,5} &= \II_9(0,0,1,0,1,1,0,1,1,1,2,0). \qquad  &\II_{9,6} = \II_9(1,0,1,0,1,1,0,1,1,1,1,0). \\
\II_{9,7} &= \II_9(1,0,1,1,1,1,0,1,1,1,1,0). \qquad  &\II_{9,8} = \II_9(1,0,1,1,1,1,0,1,1,1,2,0). \\
\II_{9,9} &= \II_9(1,0,1,1,1,1,0,1,1,2,1,0). \qquad  &\II_{9,10} = \II_9(0,0,1,0,1,1,1,1,0,1,0,0). \\
\II_{9,11} &= \II_9(0,0,1,1,1,1,1,0,1,1,0,0). \qquad  &\II_{9,12} = \II_9(0,0,1,1,1,0,0,1,1,1,0,1). \\
\II_{9,13} &= \II_9(0,0,1,1,1,1,0,1,1,1,0,1). \qquad  &\II_{9,14} = \II_9(0,0,1,1,1,1,0,1,1,1,0,2). \\
\II_{9,15} &= \II_9(0,0,1,1,1,1,1,1,1,1,0,1). \qquad  &\II_{9,16} = \II_9(0,0,1,1,1,1,2,1,1,1,0,1). \\
\II_{9,17} &= \II_9(0,0,1,1,1,1,1,1,1,1,0,2). \qquad  &\II_{9,18} = \II_9(1,0,1,0,1,0,0,1,1,1,1,1). \\
\II_{9,19} &= \II_9(1,0,1,0,1,1,0,1,1,1,1,1). \qquad  &
\end{align*}
The MIs belonging to the remaining integral families ($\II_{11} - \II_{24}$) are
\begin{align*}
\II_{11,1} &= \II_{11}(1,0,1,1,0,1,0,0,1,1,1,0). \qquad  &\II_{11,2} = \II_{11}(1,0,1,1,1,1,0,1,1,1,1,0). \\
\II_{11,3} &= \II_{11}(1,0,1,1,1,1,0,1,1,1,2,0). \qquad  &\II_{11,4} = \II_{11}(1,0,1,1,1,1,0,1,1,2,1,0). \\
\II_{11,5} &= \II_{11}(1,0,1,1,0,1,0,1,1,1,1,0). \qquad  &\II_{13,1} = \II_{13}(1,0,0,0,1,1,0,1,1,0,1,1). \\
\II_{13,2} &= \II_{13}(1,0,0,0,1,1,0,1,1,0,1,2). \qquad  &\II_{13,3} = \II_{13}(1,0,1,0,1,0,0,1,1,1,1,1). \\
\II_{13,4} &= \II_{13}(0,0,1,0,1,1,0,1,1,1,1,1). \qquad  &\II_{13,5} = \II_{13}(1,0,0,0,1,1,0,1,1,1,1,1). \\
\II_{13,6} &= \II_{13}(1,0,0,0,1,1,0,1,1,1,1,2). \qquad  &\II_{14,1} = \II_{14}(1,1,0,1,1,1,0,0,1,1,0,1). \\
\II_{14,2} &= \II_{14}(1,1,0,1,1,1,0,1,1,1,0,1). \qquad  &\II_{14,3} = \II_{14}(1,1,0,1,1,1,0,1,1,1,0,2). \\
\II_{16,1} &= \II_{16}(1,0,1,1,1,1,0,1,1,1,0,0). \qquad  &\II_{16,2} = \II_{16}(0,0,1,1,1,1,1,1,1,1,0,0). \\
\II_{16,3} &= \II_{16}(1,0,1,1,1,1,1,1,1,1,0,0). \qquad  &\II_{17,1} = \II_{17}(1,0,1,1,1,0,1,1,1,0,0,1). \\
\II_{17,2} &= \II_{17}(1,0,1,1,1,0,1,1,1,0,0,2). \qquad  &\II_{17,3} = \II_{17}(1,0,1,1,1,1,1,1,1,0,0,1). \\
\II_{17,4} &= \II_{17}(1,0,1,1,1,2,1,1,1,0,0,1). \qquad  &\II_{17,5} = \II_{17}(1,0,1,1,1,1,1,1,1,0,0,2). \\
\II_{18,1} &= \II_{18}(0,0,1,1,1,1,0,1,1,1,1,0). \qquad  &\II_{18,2} = \II_{18}(0,0,0,1,1,1,0,1,1,0,1,1). \\
\II_{18,3} &= \II_{18}(0,0,1,1,1,1,0,1,1,0,1,1). \qquad  &\II_{18,4} = \II_{18}(1,0,0,1,1,1,0,1,1,1,1,0). \\
\II_{18,5} &= \II_{18}(2,0,0,1,1,1,0,1,1,1,1,0). \qquad  &\II_{18,6} = \II_{18}(1,0,1,1,1,1,0,0,1,1,1,0). \\
\II_{18,7} &= \II_{18}(2,0,1,1,1,1,0,0,1,1,1,0). \qquad  &\II_{18,8} = \II_{18}(1,0,1,1,1,1,0,1,1,1,1,0). \\
\II_{18,9} &= \II_{18}(2,0,1,1,1,1,0,1,1,1,1,0). \qquad  &\II_{20,1} = \II_{20}(1,1,0,1,1,1,0,1,1,1,0,1). \\
\II_{20,2} &= \II_{20}(1,1,0,1,1,1,0,1,1,1,0,2). \qquad  &\II_{20,3} = \II_{20}(1,1,0,1,1,1,0,1,1,2,0,1). \\
\II_{21,1} &= \II_{21}(1,1,0,1,1,1,0,0,0,1,1,1). \qquad  &\II_{23,1} = \II_{23}(1,1,0,1,1,1,0,0,0,1,1,1). \\
\II_{23,2} &= \II_{23}(1,1,0,1,1,1,1,0,0,1,1,1). \qquad  &\II_{24,1} = \II_{24}(1,1,0,1,1,1,0,0,0,1,1,1). \\
\end{align*}
In Fig.~\ref{fig:mis}, we present schematic representations of the appearing nine-propagator MIs with the corresponding sector, for each integral families. 
MIs found in sub-sectors are obtained by pinching one or more propagators.

\subsection{Computation of the master integrals}
We employ the method of differential equations \cite{Kotikov:1990kg,Remiddi:1997ny,Gehrmann:1999as,Argeri:2007up,Henn:2013pwa,Henn:2014qga,Ablinger:2015tua,Ablinger:2018zwz} to compute the MIs. This involves differentiating the MIs with respect to $x$, followed by the application of the 
IBP identities to reduce the resulting expressions and derive a system of first-order coupled differential equations. 
The system can be represented as
\begin{equation}
    \frac{\partial}{\partial x} \II_i=\sum_{i=1}^n  \mathcal{A}_{ik} \II_k \,,
\end{equation}
where $\II_i$ denotes the $i^{th}$ MI in a set of $n$ MIs. 
The $n \times n$ matrix $\mathcal{A}$ is the connection matrix.
The system can be strategically organized into a block-triangular form which enables a solution using either a bottom-up or top-down approach. 
These systems can also be reduced to a canonical form or $\ep$-form~\cite{Henn:2013pwa,Henn:2014qga}, which occurs when the right-hand side of the system of differential equations is proportional to $\ep$.
As we expand the system in an $\ep$-series,
this proportionality to $\ep$ enables a straightforward solution of the system in terms of iterated integrals.
General algorithms exist for finding the $\ep$-form of systems of differential equations with a single 
variable~\cite{Lee:2014ioa} and with multiple scales \cite{Meyer:2016slj}.
These algorithms have been successfully implemented in public codes such as \textsc{Fuchsia}~\cite{Gituliar:2017vzm}, 
\textsc{epsilon}~\cite{Prausa:2017ltv}, \textsc{Libra}~\cite{Lee:2020zfb} and \textsc{CANONICA}~\cite{Meyer:2017joq}.
Nevertheless, as exemplified in \cite{davies2018double}, determining the canonical basis across all sectors of an integral family can be challenging.
In this work, as will be explained later, several square roots arise that cannot be simultaneously rationalized by a single transformation. We believe that adopting a canonical basis would further complicate this situation. Therefore, instead of insisting on finding a canonical basis, we choose to solve the first-order coupled differential equations by decoupling the subsystems into higher-order differential equations, which are then solved using the method of variation of constants.

To summarize, we subsequently arrange the system in an upper block-triangular structure and solve it iteratively using a bottom-up approach.
In this approach, the final block (representing a coupled subsystem) is homogeneously coupled, allowing for a bottom-up calculation. We expand each block in a series in $\ep$ and solve iteratively for each order in $\ep$, 
starting from the leading singular term. Furthermore, at each order of the $\ep$-expansion, the block decouples, 
resulting in higher-order differential equations. 
However, the associated operators for these higher-order differential equations are found to factorize into first-order.
Consequently, the relevant function space for this calculation is spanned exclusively by multiple polylogarithms
i.e.~the usual HPLs \cite{Remiddi:1999ew} and 
GPLs \cite{Ablinger:2013cf}. 
The complete system is then solved using the method of variation of constants.
Throughout several intermediate steps of the calculation, we have used 
\textsc{HarmonicSums} \cite{Ablinger:2010kw,Ablinger:2011te,Ablinger:2014rba}
and 
\textsc{PolyLogTools} \cite{Duhr:2019tlz}.

The complexity stemming from the integral topologies leads to the appearance of several square roots in our computation. 
We encounter the following square roots in the solution to the homogeneous equation: 
$\sqrt{(4-x)x}$,  $\sqrt{1-4x}$ and $x^{-3/2} \tan ^{-1}\left(\sqrt{x}\right)$.
To address this, we employ variable transformations, as detailed in Section \ref{sec:notation}, to rationalize these square roots. Since a single transformation cannot rationalize all of them simultaneously, we apply different transformations for different cases. Consequently, for some MIs, the non-homogeneous part contains a mixture of GPLs with two distinct arguments. In these instances, we integrate each part separately using the appropriate integration measure and determine the integration constants after integrating all components.
As anticipated, all topologically planar integral families have been solved in terms of the variable $x$.
Therefore, the final expressions for our results involve GPLs with arguments $x, x_l, x_n$ and $x_i$, 
as well as polynomials of these variables.
In the following, we present the alphabets that govern the structure of the GPLs with each distinct argument.
\begin{align}
  & G[\_\_,x] :  \{ -1,0,1\} \nonumber\\
  & G[\_\_,x_l] :  \{ -1,0,1,r_3,r_4\} \nonumber\\
  & G[\_\_,x_n] :  \{ -1,0,1,r_3,r_4,w_3,w_4\} \nonumber\\  
  & G[\_\_,x_i] :  \{ -1,0,1,i,-i\} \nonumber
\end{align}
where 
\begin{equation}
    r_3 = - \frac{1}{2} + \frac{\sqrt{3}}{2} i \,, \quad
    r_4 = - \frac{1}{2} - \frac{\sqrt{3}}{2} i \,, \quad
    w_3 = \frac{-3+\sqrt{5}}{2} \,, \quad
    w_4 = \frac{-3-\sqrt{5}}{2} \,.
\end{equation}
One challenging aspect of this calculation arises
when a MI relies on preceding integrals expressed in terms of $x$, while its homogeneous part requires a base transformation, we must transform the GPLs with argument $x$ which are appearing in the particular integral part. 
Since all three transformation rules required in our computation are non-linear, the fibration of the GPLs presents a significant challenge.
While the problem for lower-weight GPLs can be circumvented by identifying the basis for the GPLs and fitting high-precision values using PSLQ algorithm~\cite{pslq:92}, some dependent integrals in our case require these transformations up to weight 7 of the GPLs. At this weight, the high dimensionality of the basis makes PSLQ computationally expensive, if not infeasible.
To address this issue, we employ a two-step approach: differentiation followed by successive integration.
The corresponding integration constants are fixed by applying boundary conditions derived from the left-hand side of our equations. 
Beyond weight six, this methodology faces significant computational bottlenecks primarily due to the exponential growth in the number of GPLs. At this stage, GPLs involving all combinations of the letters $\{1,0,-1,r_1,...,r_4,w_1,...,w_4\}$ appear.
The size of some integrands reaches tens of megabits.
Symbolically integrating expressions of this magnitude presents a truly formidable challenge.

\subsubsection{Boundary condition}
Boundary conditions are essential for obtaining unique solutions to the differential equations.
Conventionally, these conditions are obtained by evaluating the MIs at particular values of the
kinematic variable $x$ (e.g. $x = 0, 1, -1$) using Feynman parameters or the Mellin-Barnes approach,
or by requiring regularity at these points. 
To this end, we initially computed a few two-point three-loop MIs utilizing Feynman parameters, and their closed-form solutions are presented below. 
The thick line represents the massive propagator and the double line represents the double power of the propagator.
%
%
%
\[
\mathbf{B_{4,1}:}\qquad \lim_{q^2\to 0}\quad
\begin{tikzpicture}[baseline=(base)]
   \begin{feynman}
        \vertex (a);
        \vertex[left=0.6cm of a] (al);
        \vertex[right=of a] (b);
        \vertex[right=0.6cm of b] (cr);

    \diagram* {
      (a) -- [plain,half left, looseness=1.6] (b),
      (a) -- [plain, half left, looseness=0.9] (b),
      (a) -- [plain, half right, very thick, looseness=0.9] (b),      
      (a) -- [plain,half right, looseness=1.6] (b),
      (a)--[plain] (al),
      (b)--[plain] (cr),
    };
    \end{feynman}
  \coordinate [yshift=-2.5pt] (base) at (b);
\end{tikzpicture}
\qquad =\frac{\Gamma \left(4-\frac{3 d}{2}\right) \Gamma (3-d) \Gamma
   \left(\frac{d}{2}-1\right)^3}{\Gamma \left(\frac{d}{2}\right)}
\]
\[
\mathbf{B_{5,1}:}\quad \lim_{q^2\to 0}\quad
\scalebox{0.7}{\begin{tikzpicture}[baseline=(base)]
   \begin{feynman}
        \vertex (a);
        \vertex[left=0.6cm of a] (al);
        \vertex[right=0.6cm of a] (b);
        \vertex[right=of b] (c);
        \vertex[right=0.6cm of c] (d);
        \vertex[right=0.6cm of d] (cr);

    \diagram* {
      (a) -- [plain, very thick] (b),  
      (b) -- [plain,half right, looseness=1.2] (c),
      (b) -- [plain,half left, looseness=1.2] (c),  
      (c) -- [plain, very thick] (d), 
      (a) -- [plain,half right, looseness=1.2] (d),
      (a) -- [plain,half left, looseness=1.2] (d),
      (a)--[plain] (al),
      (d)--[plain] (cr),
      
    };
    \end{feynman}
  \coordinate [yshift=-2.5pt] (base) at (b);
\end{tikzpicture}}
\qquad = -\frac{\Gamma \left(5-\frac{3 d}{2}\right) \Gamma
   \left(2-\frac{d}{2}\right)^2 \Gamma
   \left(\frac{d}{2}-1\right)^4 \Gamma \left(\frac{3
   d}{2}-4\right)}{\Gamma (d-2)^2 \Gamma
   \left(\frac{d}{2}\right)}
\]
\[
\mathbf{B_{5,2}:}\qquad \lim_{q^2\to 0}\quad
\begin{tikzpicture}[baseline=(base)]
   \begin{feynman}
        \vertex (a);
        \vertex[left=0.6cm of a] (al);
        \vertex[right=of a] (b);
        \vertex[right=0.6cm of b] (cr);

    \diagram* {
      (a) -- [plain,half left, looseness=1.6] (b),
      (a) -- [plain,half left, looseness=1.6, yshift=-2.9pt] (b),
      (a) -- [plain, half left, looseness=0.9] (b),
      (a) -- [plain, half right, very thick, looseness=0.9] (b),      
      (a) -- [plain,half right, looseness=1.6] (b),
      (a)--[plain] (al),
      (b)--[plain] (cr),
    };
    \end{feynman}
  \coordinate [yshift=-2.5pt] (base) at (b);
\end{tikzpicture}
\qquad =-\frac{\Gamma \left(5-\frac{3 d}{2}\right) \Gamma (4-d) \Gamma
   \left(\frac{d}{2}-2\right) \Gamma
   \left(\frac{d}{2}-1\right)^2}{\Gamma \left(\frac{d}{2}\right)}
\]
\[
\mathbf{B_{4,2}:}\qquad \lim_{q^2\to -1}\quad
\scalebox{0.7}{\begin{tikzpicture}[baseline=(base)]
   \begin{feynman}
        \vertex (a);
        \vertex[left=0.6cm of a] (al);
        \vertex[right =of a] (b);
        \vertex[right =of b] (c);
        \vertex[right=0.6cm of c] (cr);

    \diagram* {
      (a) -- [plain, half left, very thick] (b),
      (a) -- [plain, half right, very thick] (b),
      (b) -- [plain, half left] (c),
      (b) -- [plain, half right] (c),
      (b) -- [plain] (c),
      (a)--[plain] (al),
      (c)--[plain] (cr),
    };
    \end{feynman}
  \coordinate [yshift=-2.5pt] (base) at (a);
\end{tikzpicture}}
\qquad =\frac{\Gamma (3-d) \Gamma \left(1-\frac{d}{2}\right) \Gamma
   \left(\frac{d}{2}-1\right)^3}{\Gamma \left(\frac{3
   d}{2}-3\right)}
\]
\[
\mathbf{B_{5,3}:}\qquad \lim_{q^2\to 0}\quad
\scalebox{0.7}{\begin{tikzpicture}[baseline=(base)]
   \begin{feynman}
        \vertex (a);
        \vertex[left=0.6cm of a] (al);
        \vertex[above right =of a] (b);
        \vertex[below right =of a] (b1);
        \vertex[above right =of b1] (c);
        \vertex[right=0.6cm of c] (cr);

    \diagram* {
      (a) -- [plain, quarter left] (b) --[plain,quarter left] (c),
      (a) -- [plain, quarter right] (b1) --[plain,quarter right] (c),
      (b) --[plain,quarter left,very thick, looseness=1.2] (a),
      (b) --[plain,quarter right, looseness=1.2] (c),
      (a)--[plain] (al),
      (c)--[plain] (cr),
    };
    \end{feynman}
  \coordinate [yshift=-2.5pt] (base) at (a);
\end{tikzpicture}}
\; =-\frac{\Gamma \left(5-\frac{3 d}{2}\right) \Gamma (4-d) \Gamma
   \left(2-\frac{d}{2}\right) \Gamma \left(\frac{d}{2}-1\right)^3
   \Gamma (d-3)}{\Gamma \left(3-\frac{d}{2}\right) \Gamma (d-2)
   \Gamma \left(\frac{d}{2}\right)}
\]
\[
\mathbf{B_{4,3}:}\qquad \lim_{q^2\to -1}\quad
\begin{tikzpicture}[baseline=(base)]
   \begin{feynman}
        \vertex (a);
        \vertex[left=0.6cm of a] (al);
        \vertex[right=of a] (b);
        \vertex[right=0.6cm of b] (cr);

    \diagram* {
      (a) -- [plain,half left, looseness=1.6] (b),
      (a) -- [plain, half left, looseness=0.9] (b),
      (a) -- [plain, half right, looseness=0.9] (b),      
      (a) -- [plain,half right, looseness=1.6] (b),
      (a)--[plain] (al),
      (b)--[plain] (cr),
    };
    \end{feynman}
  \coordinate [yshift=-2.5pt] (base) at (b);
\end{tikzpicture}
\qquad =\frac{\Gamma \left(4-\frac{3 d}{2}\right) \Gamma
   \left(\frac{d}{2}-1\right)^4}{\Gamma (2 d-4)}
\]
\[
\mathbf{B_{5,4}:}\qquad \lim_{q^2\to -1}\quad
\scalebox{0.7}{\begin{tikzpicture}[baseline=(base)]
   \begin{feynman}
        \vertex (a);
        \vertex[left =0.6cm of a] (al);
        \vertex[right =of a] (b);
        \vertex[right =of b] (c);
        \vertex[right=0.6cm of c] (cr);

    \diagram* {
      (a) -- [plain, half left] (b),
      (a) -- [plain, half right] (b),
      (b) -- [plain, half left] (c),
      (b) -- [plain, half right] (c),
      (b) -- [plain,very thick] (c),
      (a)--[plain] (al),
      (c)--[plain] (cr),
    };
    \end{feynman}
  \coordinate [yshift=-2.5pt] (base) at (a);
\end{tikzpicture}}
\qquad =-\frac{\Gamma (3-d) \Gamma \left(2-\frac{d}{2}\right)^2 \Gamma
   \left(\frac{d}{2}-1\right)^4}{\Gamma (d-2) \Gamma
   \left(\frac{d}{2}\right)}
\]
\[
\mathbf{B_{5,5}:}\; \lim_{q^2\to 0}\;
\scalebox{0.5}{\begin{tikzpicture}[baseline=(base)]
   \begin{feynman}
        \vertex (a);
        \vertex[right=0.6cm of a] (b);
        \vertex[right=of b] (c);
        \vertex[right=0.6cm of c] (d);

    \diagram* {
      (a) -- [plain, very thick] (b),  
      (b) -- [plain,half right, looseness=1.2] (c),
      (b) -- [plain,half left, looseness=1.2] (c),  
      (c) -- [plain, very thick] (d), 
      (a) -- [plain,half right, looseness=1.2] (d),
      (a) -- [plain,half left, looseness=1.2] (d),
      (a) -- [plain,half left, looseness=1.2, yshift=-2.9pt] (d),
      
    };
    \end{feynman}
  \coordinate [yshift=-2.5pt] (base) at (b);
\end{tikzpicture}}
\quad =\frac{\Gamma \left(6-\frac{3 d}{2}\right) \Gamma
   \left(2-\frac{d}{2}\right) \Gamma \left(3-\frac{d}{2}\right)
   \Gamma \left(\frac{d}{2}-2\right) \Gamma
   \left(\frac{d}{2}-1\right)^3 \Gamma \left(\frac{3
   d}{2}-5\right)}{\Gamma (d-3) \Gamma (d-2) \Gamma
   \left(\frac{d}{2}\right)}
\]
\[
\mathbf{T_{6,1}:}\; \lim_{q^2\to 0}\;
    \scalebox{0.4}{
\begin{tikzpicture}[baseline=(base)]
    \begin{feynman}
        \vertex (a);
        \vertex[below right=1.0cm of a] (b);
        \vertex[below right=1.3cm of b] (c);
        \vertex[below right=1.3cm of c] (d);
        \vertex[right= 2.0cm of d] (e);
        \vertex[below left=1.3cm of d] (f);
        \vertex[below left=1.3cm of f] (g);
        \vertex[below left=1.0cm of g] (h);
    \diagram*{
    (a)-- [plain, edge label=$p_1$] (b),
    (b)-- [plain] (c),
    (c)--[plain,quarter right,looseness=] (d),
    (c)-- [plain] (d),
    (d)-- [plain, edge label=$q$] (e),
    (d)--[plain] (f),
    (f)--[plain] (g),
    (g)--[plain, edge label=$p_2$] (h),
    (c)--[plain, very thick] (f),
    (c)--[plain] (g),
    };
        
    \end{feynman}
      \coordinate [yshift=-2.5pt] (base) at (e);
\end{tikzpicture}
}
\quad =\frac{\Gamma \left(6-\frac{3 d}{2}\right) \Gamma (5-d) \Gamma
   \left(2-\frac{d}{2}\right) \Gamma \left(\frac{d}{2}-2\right)
   \Gamma \left(\frac{d}{2}-1\right)^2 \Gamma (d-3)}{\Gamma
   \left(3-\frac{d}{2}\right) \Gamma (d-2) \Gamma
   \left(\frac{d}{2}\right)}
\]
\[
\mathbf{B_{5,5}:}\qquad \lim_{q^2\to -1}\quad
\scalebox{0.7}{\begin{tikzpicture}[baseline=(base)]
   \begin{feynman}
        \vertex (a);
        \vertex[left =0.6cm of a] (al);
        \vertex[right =of a] (b);
        \vertex[right =of b] (c);
        \vertex[right=0.6cm of c] (cr);

    \diagram* {
      (a) -- [plain, half left] (b),
      (a) -- [plain, half right] (b),
      (b) -- [plain, half left] (c),
      (b) -- [plain, half right] (c),
      (b) -- [plain] (c),
      (a)--[plain] (al),
      (c)--[plain] (cr),
    };
    \end{feynman}
  \coordinate [yshift=-2.5pt] (base) at (a);
\end{tikzpicture}}
\qquad =-\frac{\Gamma (3-d) \Gamma \left(2-\frac{d}{2}\right) \Gamma
   \left(\frac{d}{2}-1\right)^5}{\Gamma (d-2) \Gamma
   \left(\frac{3 d}{2}-3\right)}
\]
\[
\mathbf{B_{5,6}:}\qquad \lim_{q^2\to -1}\quad
\scalebox{0.7}{\begin{tikzpicture}[baseline=(base)]
   \begin{feynman}
        \vertex (a);
        \vertex[left=0.6cm of a] (al);
        \vertex[above right =of a] (b);
        \vertex[below right =of a] (b1);
        \vertex[above right =of b1] (c);
        \vertex[right=0.6cm of c] (cr);

    \diagram* {
      (a) -- [plain, quarter left] (b) --[plain,quarter left] (c),
      (a) -- [plain, quarter right] (b1) --[plain,quarter right] (c),
      (b) --[plain,quarter left,looseness=1.2] (a),
      (b) --[plain,quarter right, looseness=1.2] (c),
      (a)--[plain] (al),
      (c)--[plain] (cr),
    };
    \end{feynman}
  \coordinate [yshift=-2.5pt] (base) at (a);
\end{tikzpicture}}
\qquad =-\frac{\Gamma \left(5-\frac{3 d}{2}\right) \Gamma
   \left(2-\frac{d}{2}\right)^2 \Gamma
   \left(\frac{d}{2}-1\right)^5 \Gamma \left(\frac{3
   d}{2}-4\right)}{\Gamma (4-d) \Gamma (d-2)^2 \Gamma (2 d-5)}
\]
\[
\mathbf{B_{6,1}:}\qquad \lim_{q^2\to -1}\quad
\scalebox{0.7}{\begin{tikzpicture}[baseline=(base)]
   \begin{feynman}
        \vertex (a);
        \vertex[left=0.6cm of a] (al);
        \vertex[right =of a] (b);
        \vertex[right =of b] (c);
        \vertex[right =of c] (d);
        \vertex[right=0.6cm of d] (cr);

    \diagram* {
      (a) -- [plain, half left] (b),
      (a) -- [plain, half right] (b),
      (b) -- [plain, half left] (c),
      (b) -- [plain, half right] (c),
      (c) -- [plain, half left] (d),
      (c) -- [plain, half right] (d),
      (a)--[plain] (al),
      (d)--[plain] (cr),
    };
    \end{feynman}
  \coordinate [yshift=-2.5pt] (base) at (a);
\end{tikzpicture}}
\qquad =\frac{\Gamma \left(2-\frac{d}{2}\right)^3 \Gamma
   \left(\frac{d}{2}-1\right)^6}{\Gamma (d-2)^3}
\]
\[
\mathbf{T_{5,1}:}\quad \lim_{q^2\to 0}\;
\scalebox{0.6}{
\begin{tikzpicture}[baseline=(base)]
    \begin{feynman}
        \vertex (a);
        \vertex[below right=1.3cm of a] (b);
        \vertex[below right=1.3cm of b] (c);
        \vertex[right= 2.0cm of c] (d);
        \vertex[below left=1.3cm of c] (e);
        \vertex[below left=1.3cm of e] (f);
    \diagram*{
    (a)-- [plain] (b),
    (b)-- [plain] (c),
    (b)--[plain, quarter right, looseness=1.2, very thick] (c),
    (c)-- [plain] (d),
    (c)-- [plain] (e),
    (e)-- [plain] (f),
    (e)--[plain, quarter left, looseness=1.2] (b),
    (b)-- [plain] (e),
    };
        
    \end{feynman}
 \coordinate [yshift=-2.5pt] (base) at (d);
\end{tikzpicture}
}
\quad =-\frac{\Gamma \left(5-\frac{3 d}{2}\right) \Gamma (4-d) \Gamma
   \left(2-\frac{d}{2}\right) \Gamma \left(\frac{d}{2}-1\right)^3
   \Gamma (d-3)}{\Gamma \left(3-\frac{d}{2}\right) \Gamma (d-2)
   \Gamma \left(\frac{d}{2}\right)}
\]
\[
\mathbf{T_{5,2}:}\quad \lim_{q^2\to 0}\;
\scalebox{0.6}{
\begin{tikzpicture}[baseline=(base)]
    \begin{feynman}
        \vertex (a);
        \vertex[below right=1.3cm of a] (b);
        \vertex[below right=1.3cm of b] (c);
        \vertex[right= 2.0cm of c] (d);
        \vertex[below left=1.3cm of c] (e);
        \vertex[below left=1.3cm of e] (f);
    \diagram*{
    (a)-- [plain] (b),
    (b)-- [plain] (c),
    (b)--[plain, quarter right, looseness=1.2, very thick] (c),
    (c)-- [plain] (d),
    (c)-- [plain] (e),
    (c)-- [plain, yshift=2.9pt] (e),
    (e)-- [plain] (f),
    (e)--[plain, quarter left, looseness=1.2] (b),
    (b)-- [plain] (e),
    };
        
    \end{feynman}
 \coordinate [yshift=-2.5pt] (base) at (d);
\end{tikzpicture}
}
\quad =\frac{\Gamma \left(6-\frac{3 d}{2}\right) \Gamma (5-d) \Gamma
   \left(2-\frac{d}{2}\right) \Gamma \left(\frac{d}{2}-1\right)^3
   \Gamma (d-4)}{\Gamma \left(4-\frac{d}{2}\right) \Gamma (d-2)
   \Gamma \left(\frac{d}{2}\right)}
\]
\[
\mathbf{B_{5,7}:}\qquad \lim_{q^2\to 0}\quad
\scalebox{0.7}{\begin{tikzpicture}[baseline=(base)]
   \begin{feynman}
        \vertex (a);
        \vertex[left=0.6cm of a] (al);
        \vertex[above right =of a] (b);
        \vertex[below right =of a] (b1);
        \vertex[above right =of b1] (c);
        \vertex[right=0.6cm of c] (cr);

    \diagram* {
      (a) -- [plain, quarter left] (b) --[plain,quarter left] (c),
      (a) -- [plain, quarter right, very thick] (b1) --[plain,quarter right, very thick] (c),
      (b) --[plain,quarter left,looseness=1.2] (a),
      (b) --[plain,quarter right, looseness=1.2] (c),
      (a)--[plain] (al),
      (c)--[plain] (cr),
    };
    \end{feynman}
  \coordinate [yshift=-2.5pt] (base) at (a);
\end{tikzpicture}}
\qquad =-\frac{\Gamma \left(5-\frac{3 d}{2}\right) \Gamma
   \left(2-\frac{d}{2}\right)^2 \Gamma
   \left(\frac{d}{2}-1\right)^4 \Gamma \left(\frac{3
   d}{2}-4\right)}{\Gamma (d-2)^2 \Gamma
   \left(\frac{d}{2}\right)}
\]
\[
\mathbf{B_{5,8}:}\; \lim_{q^2\to 0}\;
\scalebox{0.7}{\begin{tikzpicture}[baseline=(base)]
   \begin{feynman}
        \vertex (a);
        \vertex[above right =of a] (b);
        \vertex[below right =of a] (b1);
        \vertex[above right =of b1] (c);

    \diagram* {
      (a) -- [plain, quarter left] (b) --[plain,quarter left] (c),
      (a) -- [plain, quarter right, very thick] (b1) --[plain,quarter right, very thick] (c),
      (b) --[plain,quarter left,looseness=1.2] (a),
      (a) --[plain,quarter left,yshift=-4.9pt] (b),
      (b) --[plain,quarter right, looseness=1.2] (c),
    };
    \end{feynman}
  \coordinate [yshift=-2.5pt] (base) at (a);
\end{tikzpicture}}
\; =\frac{\Gamma \left(6-\frac{3 d}{2}\right) \Gamma
   \left(2-\frac{d}{2}\right) \Gamma \left(3-\frac{d}{2}\right)
   \Gamma \left(\frac{d}{2}-2\right) \Gamma
   \left(\frac{d}{2}-1\right)^3 \Gamma \left(\frac{3
   d}{2}-5\right)}{\Gamma (d-3) \Gamma (d-2) \Gamma
   \left(\frac{d}{2}\right)}
\]
\[
\mathbf{B_{5,9}:}\qquad \lim_{q^2\to -1}\quad
\scalebox{0.7}{\begin{tikzpicture}[baseline=(base)]
   \begin{feynman}
        \vertex (a);
        \vertex[right =0.6cm of a] (b);
        \vertex[right =of b] (c);
        \vertex[right =of c] (d);
        \vertex[right =0.6cm of d] (e);

    \diagram* {
      (a) -- [plain] (b),
      (b) -- [plain, half left] (c),
      (b) -- [plain, half right] (c),
      (c) -- [plain, half left] (d),
      c --[plain, out=118, in=62, loop, min distance=2cm, very thick] c,
      (c) -- [plain, half right] (d),
      (d) --[plain] (e),
    };
    \end{feynman}
  \coordinate [yshift=-2.5pt] (base) at (a);
\end{tikzpicture}}
\qquad =-\frac{\Gamma \left(1-\frac{d}{2}\right) \Gamma
   \left(2-\frac{d}{2}\right)^2 \Gamma
   \left(\frac{d}{2}-1\right)^4}{\Gamma (d-2)^2}
\]
%
%
%
For the majority of topologically planar MIs, the boundary conditions can be determined using the 
aforementioned results or by imposing regularity conditions at specific values of $x$.
However, for more complex cases, this procedure becomes cumbersome and impractical.
To automate the determination of these boundary constants, we turn to the auxiliary mass flow
method \cite{Liu:2017jxz,Liu:2018dmc,Liu:2021wks} as implemented in \textsc{AMFlow}. This allows us to obtain highly precise numerical values for the MIs at particular kinematic points, which can then be used
with the \textsc{PSLQ} algorithm \cite{pslq:92} to reconstruct analytic expressions,
provided we can anticipate the relevant set of constants.
The study of differential equations and corresponding polylogarithms suggests that 
the appearing constants are primarily multiple zeta values (MZVs) \cite{Blumlein:2009cf}. 
However, depending on the chosen kinematic point, the MIs may also include constants such as
$\ln (2)$, ${{\rm Li}_n ({\scriptstyle \frac{1}{2}})}$ and other cyclotomic constants~\cite{Ablinger:2011te}.
To fix boundary conditions, we need to expand GPLs around chosen kinematic points, typically 
$x = 0, 1$ or $-1$.
While this expansion yields constants like MZVs,
$\ln (2)$, or ${{\rm Li}_n ({\scriptstyle \frac{1}{2}})}$
for HPLs, it becomes significantly more complex for GPLs involving the alphabets 
$\{r_3, r_4\}$, or $\{w_3, w_4\}$ or both.
We use the tables provided in Ref.~\cite{Henn:2015sem} for 
the set of letters $\{r_3, r_4\}$ up to weight 6. 
A study to establish the basis of constant GPLs involving the alphabet $\{w_1,...,w_4 \}$ is planned for future investigation.

\section{Results}
\label{sec:result}
In this section, we present our findings, which include the computation of three-point 
three-loop MIs with one internal massive propagator. These MIs, in conjunction with those 
presented in \cite{Bonetti:2017ovy}, will enable the determination of three-loop mixed 
QCD-EW (${\mathcal O} (\alpha \alpha_s^2)$) corrections to the quark form factor.

Due to their immense size, the complete analytic expressions for the MIs cannot be presented within this paper. The ancillary files containing these expressions are available in the arXiv version of this paper.
The results have been presented in a \textsc{Mathematica} replaceable list format, where the left side for each element of the list contains the MIs already presented before in \textsc{LiteRed} notation \texttt{j[family,indices][variable]}.  On the right side, the analytical results are provided in terms of the corresponding kinematic variables. It is important to note that there are slight differences in the naming conventions employed in this paper compared to those used in the file, as follows
\begin{equation}
    x_l : \text{xL} \,, \quad 
    x_n : \text{xN} \,, \quad 
    x_i : \text{xx} \,.    
\end{equation}

To illustrate our results, we present the numerical evaluation of each MI 
at $x=\frac{1}{11}$ below.
Each MI, denoted as $\II_{m,n}$, has been multiplied by a factor of $\ep^k$ 
such that the coefficient of $\ep^6$ contains GPLs or MZVs of weight 6. All the integrals have been evaluated using the computer algebra system \texttt{GiNac}~\cite{Bauer:2000cp}.

%
%
{\footnotesize
\begin{longtable}{cccccccc}
                    & $\ep^0$  & $\ep^1$ & $\ep^2$ & $\ep^3$ & $\ep^4$ & $\ep^5$ & $\ep^6$ \\ 
                    \hline\\[-1.6ex]
 $\ep^3  \II_{1,1}$ & 0.333333 & 1.66667 & 9.77900 & 34.8916 & 140.436 & 433.090 & 1517.54 \\
 $\ep^3  \II_{1,2}$ & 0.166667 & 1.03030 & 6.52054 & 27.1526 & 114.772 & 393.362 & 1393.12 \\
 $\ep^3  \II_{1,3}$ & -0.666667 & -4.68709 & -24.0189 & -74.8749 & -214.414 & -414.453 & -911.297 \\
 $\ep^3  \II_{1,4}$ & 0.5 & 3.69895 & 17.9884 & 65.4246 & 202.310 & 549.393 & 1373.07 \\
 $\ep^3  \II_{1,5}$ & 0.333333 & 1.68182 & 9.83135 & 35.2046 & 141.237 & 436.371 & 1524.29 \\
 $\ep^3  \II_{1,6}$ & -0.333333 & -0.955863 & -6.07779 & -12.5228 & -55.2433 & -77.5210 & -345.298 \\
 $\ep^3  \II_{1,7}$ & 0.166667 & 1.01515 & 6.48334 & 26.8797 & 114.367 & 391.023 & 1391.71 \\
 $\ep^3  \II_{1,8}$ & -0.666667 & -4.73123 & -24.3198 & -76.4855 & -219.911 & -431.455 & -950.398 \\
 $\ep^2  \II_{1,9}$ & -0.00378788 & -0.0632337 & -0.605556 & -4.23554 & -24.0994 & -117.974 & -515.887 \\
 $\ep^2  \II_{1,10}$ & 0.0151515 & 0.260510 & 2.48907 & 17.1509 & 95.2667 & 453.091 & 1919.86 \\
 $\ep^2  \II_{1,11}$ & 0.0227273 & 0.356675 & 2.96144 & 17.1901 & 78.2070 & 297.021 & 980.771 \\
 $\ep^4  \II_{1,12}$ & 0 & 0.166667 & 2.69895 & 11.6444 & 50.8462 & 136.854 & 423.628 \\
 $\ep^3  \II_{1,13}$ & 5.5 & 29.6884 & 118.495 & 336.456 & 867.736 & 1961.46 & 4674.66 \\
 $\ep^3  \II_{1,14}$ & 0.333333 & 4.73123 & 34.9210 & 180.855 & 731.342 & 2463.99 & 7164.25 \\
 $\ep^2  \II_{1,15}$ & 0.0303030 & 0.521021 & 4.82861 & 31.7308 & 166.090 & 739.024 & 2919.65 \\
 $\ep^4  \II_{1,16}$ & 0 & 0 & 0 & -14.2416 & -96.1451 & -509.299 & -1828.75 \\
 $\ep^3  \II_{1,17}$ & 1. & 9.79579 & 51.6563 & 192.344 & 564.985 & 1390.76 & 2981.75 \\
 $\ep^5  \II_{1,18}$ & 0 & 0 & 0 & 0 & 0 & -35.7867 & -258.020 \\
 $\ep^5  \II_{1,19}$ & 0 & 0 & 0 & 3.48915 & 29.4281 & 185.875 & 809.128 \\
 $\ep^3  \II_{1,20}$ & 1. & 13.1937 & 90.5693 & 428.972 & 1572.25 & 4747.58 & 12289.6 \\
 $\ep^4  \II_{1,21}$ & 0 & 0 & 0 & -14.2416 & -128.449 & -666.524 & -2485.93 \\
 $\ep^5  \II_{1,22}$ & 0 & 0 & 0 & 3.48915 & 25.9149 & 116.521 & 378.645 \\
 $\ep^4  \II_{1,23}$ & 0 & -0.333333 & -0.977717 & -6.16913 & -11.0592 & -51.1596 & -46.6273 \\
 $\ep^3  \II_{1,24}$ & 0.5 & 5.39790 & 33.8710 & 156.548 & 593.795 & 1954.57 & 5798.69 \\
 $\ep^3  \II_{1,25}$ & 0.5 & 7.09684 & 55.0264 & 305.005 & 1350.83 & 5080.83 & 16880.9 \\
 $\ep^6  \II_{1,26}$ & 0 & 0 & 0 & 0 & 0 & 135.812 & 1805.97 \\
 $\ep^5  \II_{1,27}$ & 0 & 0 & 0 & 3.48915 & 23.0060 & 138.320 & 579.505 \\
 $\ep^3  \II_{1,28}$ & 0.333333 & 3.03228 & 13.7580 & 52.7334 & 146.978 & 403.211 & 808.705 \\
 $\ep^4  \II_{1,29}$ & 0 & 0 & 0 & 2.40411 & 9.76114 & 62.6288 & 178.379 \\
 $\ep^5  \II_{1,30}$ & 0 & 0 & 0 & 0 & 13.7344 & 108.451 & 634.930 \\
 $\ep^5  \II_{1,31}$ & 0 & 0 & 0 & -2.23904 & -14.6695 & -90.3170 & -376.098 \\
 $\ep^4  \II_{1,32}$ & 0 & 0 & -2.21009 & -14.3541 & -88.2894 & -366.399 & -1478.54 \\
 $\ep^4  \II_{1,33}$ & 0 & 0 & 0 & 13.8590 & 109.949 & 645.021 & 2734.95 \\
 $\ep^6  \II_{1,34}$ & 0 & 0 & 0 & 0 & 0 & 0 & -41.0137 \\
 $\ep^5  \II_{1,35}$ & 0 & 0 & -3.01571 & -23.1674 & -103.213 & -302.595 & -552.566 \\
 $\ep^4  \II_{1,36}$ & 0 & 0 & 0 & 2.40411 & 9.19787 & 56.9336 & 141.469 \\
 $\ep^3  \II_{1,37}$ & 0 & 0 & 12.2932 & 88.1309 & 479.094 & 1828.83 & 6413.12 \\
 $\ep^6  \II_{1,38}$ & 0 & 0 & 0 & 0 & 0 & 16.1956 & 141.713 \\
 $\ep^4  \II_{1,39}$ & 0 & 0 & -3.66667 & -0.251478 & 121.964 & 744.669 & 2267.69 \\
 $\ep^5  \II_{1,40}$ & 0 & 0 & 0 & 0 & 0 & -53.8090 & -351.245 \\
 $\ep^6  \II_{1,41}$ & 0 & 0 & 0 & 0 & 0 & 12.1626 & 62.9186 \\
 $\ep^3  \II_{1,42}$ & 0.166667 & 2.69895 & 11.5652 & 51.8737 & 136.252 & 430.716 & 797.080 \\
 $\ep^4  \II_{1,43}$ & 0 & 0.5 & 7.09684 & 41.1674 & 163.655 & 497.015 & 1257.62 \\
 $\ep^3  \II_{1,44}$ & 11. & 85.7537 & 354.712 & 992.475 & 2038.14 & 3033.02 & 2608.56 \\
 $\ep^4  \II_{1,45}$ & 0 & 0.333333 & 4.73123 & 34.9210 & 133.962 & 458.375 & 1071.56 \\
 $\ep^3  \II_{1,46}$ & 5.5 & 67.0653 & 295.841 & 897.305 & 1875.95 & 2982.43 & 2622.27 \\
 $\ep^3  \II_{1,47}$ & 0.333333 & 3.02849 & 13.7023 & 52.5646 & 146.221 & 401.967 & 803.867 \\
 $\ep^3  \II_{1,48}$ & 5.5 & 29.6051 & 115.271 & 321.555 & 772.041 & 1584.42 & 2927.14 \\
 $\ep^4  \II_{1,49}$ & 0 & -0.166667 & -2.36561 & -6.24646 & -29.7734 & -38.8746 & -187.273 \\
 $\ep^5  \II_{1,50}$ & 0 & 0 & -0.166667 & -2.36561 & -6.20672 & -29.6177 & -38.1562 \\
 $\ep^5  \II_{1,51}$ & 0 & 0 & 0.166667 & 2.69895 & 23.8584 & 106.034 & 418.428 \\
 $\ep^6  \II_{1,52}$ & 0 & 0 & 0 & 0 & 0 & 12.3415 & 64.5699 \\
 $\ep^2  \II_{1,53}$ & 0.0227273 & 0.279450 & 1.88399 & 9.07998 & 34.9803 & 114.507 & 331.522 \\
 $\ep^2  \II_{1,54}$ & -0.0757576 & -0.590680 & -3.83951 & -17.9554 & -77.8544 & -291.577 & -1058.85 \\
 $\ep^3  \II_{1,55}$ & -0.166667 & -0.681818 & -4.24488 & -12.8984 & -55.3260 & -147.820 & -562.465 \\
 $\ep^4  \II_{1,56}$ & 0 & 0.166667 & 2.69895 & 11.6444 & 52.9824 & 144.607 & 475.277 \\
 $\ep^3  \II_{1,57}$ & 5.5 & 29.6884 & 118.003 & 329.721 & 815.663 & 1664.30 & 3270.53 \\
 $\ep  \II_{1,58}$ & 0.000229568 & 0.00436800 & 0.0447797 & 0.324924 & 1.87037 & 9.09857 & 38.9918 \\
 $\ep^5  \II_{1,59}$ & 0 & 0 & 0.166667 & 2.69895 & 25.5033 & 124.703 & 540.799 \\
 $\ep^4  \II_{1,60}$ & 0 & 0 & 24.3110 & 137.627 & 511.413 & 1217.30 & 1824.90 \\
 $\ep^4  \II_{1,61}$ & 2.75 & 15.9384 & 54.3482 & 107.445 & 86.7984 & -379.069 & -2052.43 \\
 $\ep^2  \II_{1,62}$ & -0.0833333 & -0.625 & -4.04892 & -18.6400 & -80.7081 & -299.675 & -1088.50 \\
 $\ep^3  \II_{1,63}$ & 0.5 & 3.72167 & 18.1217 & 65.9682 & 203.901 & 553.377 & 1381.49 \\
 $\ep^3  \II_{1,64}$ & -0.5 & -2.65481 & -10.2897 & -28.1499 & -67.3897 & -134.185 & -250.839 \\
 $\ep^5  \II_{1,65}$ & 0 & 0 & -0.333333 & -5.39790 & -29.9593 & -139.510 & -449.225 \\
 $\ep^6  \II_{1,66}$ & 0 & 0 & 0 & 0 & 0 & 17.6366 & 123.811 \\
 $\ep^5  \II_{1,67}$ & 0 & 0 & 0 & 48.6220 & 384.531 & 1905.36 & 6814.54 \\
 $\ep^3  \II_{1,68}$ & 0.166667 & 2.69895 & 11.5652 & 52.6740 & 143.025 & 471.363 & 971.487 \\
 $\ep^4  \II_{1,69}$ & 0 & 0 & 0 & -14.2416 & -115.026 & -679.570 & -2886.52 \\
 $\ep^3  \II_{1,70}$ & 0.166667 & 1.01515 & 6.26003 & 25.4120 & 105.947 & 359.826 & 1280.43 \\
 $\ep^3  \II_{1,71}$ & -0.166667 & -0.311265 & -2.80380 & -3.11092 & -25.3234 & -9.82288 & -183.490 \\
 $\ep^6  \II_{1,72}$ & 0 & 0 & 0 & 0 & 0 & 0 & -46.3370 \\
 $\ep^5  \II_{1,73}$ & 0 & 0 & 0 & 23.8963 & 116.998 & 292.284 & -175.022 \\
 $\ep^4  \II_{1,74}$ & 1.83333 & 7.63225 & 9.18605 & -38.1567 & -154.038 & 154.833 & 3774.81 \\
 $\ep^3  \II_{1,75}$ & 0.166667 & 2.69895 & 25.5033 & 175.561 & 979.208 & 4671.78 & 19821.1 \\
 $\ep^4  \II_{1,76}$ & 0 & 0 & 0 & 2.40411 & 10.1491 & 67.3481 & 211.638 \\
 $\ep^5  \II_{1,77}$ & 0 & 0 & 0 & -2.17239 & -13.3190 & -75.3476 & -265.112 \\
 $\ep^5  \II_{1,78}$ & 0 & 0 & 0.244537 & -0.618519 & -6.53046 & -58.2403 & -257.429 \\
 $\ep^4  \II_{1,79}$ & 0 & 0 & -2.21009 & -13.6095 & -76.6512 & -268.531 & -891.978 \\
 $\ep^4  \II_{1,80}$ & 0 & 0.333333 & 5.06456 & 31.0831 & 153.548 & 570.550 & 1951.52 \\
 $\ep^3  \II_{1,81}$ & 5.5 & 42.8768 & 196.450 & 613.220 & 1415.12 & 2040.28 & -226.027 \\
 $\ep^5  \II_{1,82}$ & 0 & 0 & -0.333333 & -5.39790 & -30.0595 & -140.081 & -452.089 \\
 $\ep^6  \II_{1,83}$ & -6.72222 & -48.3576 & -244.448 & -1175.90 & -4818.86 & -14536.9 & -26241.9 \\
 $\ep^5  \II_{1,84}$ & 168.056 & -123.556 & 1118.50 & 25235.1 & 181168. & 677457. & 1.74257 10$^6$ \\
 $\ep^6  \II_{1,85}$ & 67.2222 & 416.353 & 2218.46 & 6955.84 & 7818.37 & -53829.3 & -452892. \\
 $\ep^6  \II_{1,86}$ & 0 & 0 & 0 & 0 & 0 & 13.7962 & 82.0374 \\
 $\ep^3  \II_{1,87}$ & 0.333333 & 5.06456 & 44.9421 & 295.198 & 1601.43 & 7588.58 & 32634.7 \\
 $\ep^5  \II_{1,88}$ & 0 & 0 & 0 & 0 & 0 & -61.2839 & -604.110 \\
 $\ep^4  \II_{1,89}$ & 0 & -3.66667 & -3.66667 & -34.2357 & -348.953 & -2252.29 & -8923.10 \\
 $\ep^6  \II_{1,90}$ & 0 & 0 & 0 & 0 & 0 & 19.2223 & 192.820 \\
 $\ep^6  \II_{1,91}$ & 0 & 0 & 0 & 0 & 0 & 0 & -52.9463 \\
 $\ep^5  \II_{1,92}$ & 0 & 0 & -0.978147 & -8.17179 & -67.3062 & -400.866 & -2133.15 \\
 $\ep^5  \II_{1,93}$ & 0 & 0 & -0.978147 & -6.19393 & -25.5378 & -76.2818 & -190.801 \\
 $\ep^5  \II_{1,94}$ & 13.4444 & 56.3818 & -130.591 & -1877.74 & -7452.65 & -12236.3 & 25866.9 \\
 $\ep^5  \II_{1,95}$ & -73.9444 & -787.378 & -841.709 & 17995.8 & 110453. & 293314. & 18847.3 \\
 $\ep^5  \II_{1,96}$ & 0 & 0 & 0 & 0 & 0 & -44.3973 & -339.563 \\ 
 \midrule
%
%
 $\ep^3  \II_{2,1}$ & 0 & -0.0833333 & -0.590054 & -3.61758 & -15.1497 & -59.5315 & -191.177 \\
 $\ep^4  \II_{2,2}$ & 0 & -0.166667 & -0.666667 & -4.12119 & -11.5908 & -47.0060 & -100.799 \\
 $\ep^3  \II_{2,3}$ & 0.333333 & 1.66667 & 9.59718 & 32.8349 & 124.710 & 342.859 & 1077.09 \\
 $\ep^3  \II_{2,4}$ & 0.166667 & 1.00000 & 6.04101 & 23.1200 & 89.9729 & 268.268 & 841.654 \\
 $\ep^4  \II_{2,5}$ & 0 & 0 & 1.76884 & 11.7271 & 68.7287 & 256.439 & 913.385 \\
 $\ep^3  \II_{2,6}$ & 0.333333 & 1.66667 & 9.41535 & 30.9200 & 111.487 & 276.648 & 801.947 \\
 $\ep^5  \II_{2,7}$ & 0 & 0 & 0 & 4.47808 & 40.5232 & 266.716 & 1268.65 \\
 $\ep^4  \II_{2,8}$ & 0 & 0 & -1.76884 & -20.0383 & -156.237 & -906.660 & -4474.90 \\
 $\ep^3  \II_{2,9}$ & 0.166667 & 1.00000 & 6.04101 & 22.5132 & 84.1222 & 230.315 & 665.081 \\
 $\ep^5  \II_{2,10}$ & 0 & 0 & 0 & 0 & 12.3905 & 84.8895 & 471.241 \\
 $\ep^4  \II_{2,11}$ & 0 & 0 & -1.76884 & -22.0408 & -133.443 & -608.070 & -2139.11 \\
 $\ep^4  \II_{2,12}$ & 0 & 0 & -3.53768 & -38.5414 & -249.057 & -1192.41 & -4697.33 \\
 $\ep^3  \II_{2,13}$ & 0.5 & 3.69895 & 17.8066 & 63.3134 & 188.152 & 480.115 & 1098.51 \\
 $\ep^4  \II_{2,14}$ & 0 & 0.166667 & 2.69895 & 11.6444 & 50.6580 & 135.329 & 414.685 \\
 $\ep^3  \II_{2,15}$ & 5.5 & 29.6884 & 116.495 & 313.974 & 717.621 & 1216.88 & 1619.70 \\
 $\ep^6  \II_{2,16}$ & 0 & 0 & 0 & 0 & 0 & 0 & -40.1576 \\
 $\ep^4  \II_{2,17}$ & 0 & 0 & 19.4572 & 86.8330 & 290.565 & 839.485 & 3412.57 \\
 $\ep^5  \II_{2,18}$ & 0 & 0 & 0 & 0 & -12.3905 & -96.4633 & -560.047 \\
 $\ep^5  \II_{2,19}$ & 0 & 0 & 0 & 4.47808 & 43.6054 & 255.750 & 1101.72 \\
 $\ep^3  \II_{2,20}$ & 0 & 0 & -0.0804018 & -1.12246 & -8.45312 & -45.0740 & -190.384 \\
 $\ep^6  \II_{2,21}$ & 0 & 0 & 0 & 0 & 0 & 12.9508 & 84.1352 \\
 $\ep^5  \II_{2,22}$ & 0 & 0 & 0 & 49.2589 & 398.502 & 1986.89 & 6800.72 \\
 $\ep^6  \II_{2,23}$ & 0 & 0 & 0 & 0 & 0 & 0 & -50.9871 \\
 $\ep^4  \II_{2,24}$ & 0 & -1.83333 & 1.83333 & -18.4198 & -224.859 & -1330.46 & -4152.25 \\
 $\ep^5  \II_{2,25}$ & 0 & 0 & 0 & 24.6294 & 122.423 & 317.501 & -94.7209 \\
 $\ep^5  \II_{2,26}$ & 0 & 0 & 0 & 0 & -12.3905 & -101.849 & -642.513 \\
 $\ep^6  \II_{2,27}$ & 0 & 0 & -11.0576 & -39.1425 & 20.1675 & -198.015 & -6080.56 \\
 $\ep^6  \II_{2,28}$ & 0 & 0 & 17.4393 & 1303.19 & 3886.75 & 6505.39 & 52336.1 \\
 $\ep^5  \II_{2,29}$ & 0 & 73.9444 & 546.207 & 1206.74 & -538.955 & 27774.3 & 277002. \\
 $\ep^3  \II_{2,30}$ & 0.333333 & 1.66667 & 9.79415 & 34.9723 & 140.887 & 434.633 & 1523.11 \\
 $\ep^3  \II_{2,31}$ & -0.333333 & -1.00000 & -6.26384 & -13.4862 & -57.8245 & -86.2764 & -362.103 \\
 $\ep^5  \II_{2,32}$ & 0 & 0 & 0 & 0 & 2.40411 & 9.66967 & 61.4818 \\
 $\ep^5  \II_{2,33}$ & 0 & 0 & 0 & 0 & 12.3905 & 87.0644 & 452.942 \\
 $\ep^5  \II_{2,34}$ & 0 & 0 & 0 & 0 & 0 & -6.68238 & -24.0339 \\
 $\ep^3  \II_{2,35}$ & 0.333333 & 3.03228 & 13.7655 & 52.9623 & 148.874 & 416.758 & 883.761 \\
 $\ep^3  \II_{2,36}$ & 0.166667 & 1.00000 & 6.22283 & 24.9080 & 102.659 & 333.228 & 1126.61 \\
 $\ep^4  \II_{2,37}$ & 0 & 0.166667 & 1.00000 & 6.22283 & 24.2339 & 95.7057 & 285.541 \\
 $\ep^2  \II_{2,38}$ & -0.0795455 & -0.602273 & -3.91330 & -18.1282 & -78.7549 & -294.027 & -1071.35 \\
 $\ep^4  \II_{2,39}$ & 0 & 0 & 0 & 2.40411 & 42.7196 & 422.378 & 3026.43 \\
 $\ep^6  \II_{2,40}$ & 0 & 0 & 0 & 0 & 0 & 0 & -32.8342 \\
 $\ep^5  \II_{2,41}$ & 0 & 0 & -3.01571 & -18.7599 & -48.5689 & -24.3090 & -23.0719 \\
 $\ep^4  \II_{2,42}$ & 0 & 0 & -3.53768 & -40.5439 & -246.698 & -1050.69 & -3503.19 \\
 $\ep^4  \II_{2,43}$ & 0 & 0 & -3.53768 & -36.5390 & -269.885 & -1510.80 & -7380.01 \\
 $\ep^6  \II_{2,44}$ & 0 & 0 & 0 & 0 & -5.44327 & -50.0872 & -367.407 \\
 $\ep^5  \II_{2,45}$ & 0 & 0 & 0 & 4.47808 & 43.6797 & 319.065 & 1748.12 \\
\midrule 
%
 $\ep^5  \II_{3,1}$ & 0 & 0 & 0 & 0 & -13.7344 & -140.066 & -1019.12 \\
 $\ep^4  \II_{3,2}$ & 0 & 1.75473 & 1.33449 & -75.2797 & -641.325 & -3083.09 & -10275.7 \\
 $\ep^3  \II_{3,3}$ & 0.333333 & 1.68182 & 9.81620 & 35.1252 & 140.793 & 434.870 & 1518.88 \\
 $\ep^3  \II_{3,4}$ & -0.333333 & -0.955863 & -6.25199 & -14.2737 & -67.3623 & -139.584 & -617.027 \\
 $\ep^4  \II_{3,5}$ & 0 & 0.166667 & 2.69895 & 11.5652 & 50.4756 & 135.028 & 418.530 \\
 $\ep^3  \II_{3,6}$ & 5.5 & 29.2029 & 115.103 & 320.413 & 810.359 & 1781.80 & 4163.62 \\
 $\ep^6  \II_{3,7}$ & 0 & 0 & 0 & 0 & 0 & 12.8435 & 83.2141 \\
 $\ep^5  \II_{3,8}$ & 0 & 0 & 0 & 47.7926 & 382.978 & 1894.39 & 6397.62 \\
 $\ep^6  \II_{3,9}$ & 0 & 0 & 0 & 0 & 0 & 0 & -45.1899 \\
 $\ep^5  \II_{3,10}$ & 0 & 0 & -3.01571 & -21.6941 & -97.8728 & -293.081 & -630.780 \\
 $\ep^6  \II_{3,11}$ & 0 & 0 & 0 & 0 & 0 & 0 & -48.2428 \\
 $\ep^4  \II_{3,12}$ & 0 & 0.333333 & 3.03228 & 13.7655 & 53.0889 & 150.127 & 424.330 \\
 $\ep^3  \II_{3,13}$ & 5.5 & 29.6884 & 116.003 & 328.600 & 821.670 & 1865.72 & 4258.01 \\
 $\ep^4  \II_{3,14}$ & 0 & 0.166667 & 1.00000 & 6.22283 & 24.3500 & 97.0341 & 294.702 \\
 $\ep^5  \II_{3,15}$ & 0 & 0 & 0 & 0 & 13.7344 & 103.992 & 618.604 \\
 $\ep^6  \II_{3,16}$ & 0 & 0 & 0 & 0 & 0 & 0 & -32.6436 \\
 $\ep^5  \II_{3,17}$ & 0 & 0 & -2.91384 & -16.8396 & -32.3375 & 68.2689 & 401.231 \\
 $\ep^5  \II_{3,18}$ & 0 & 0 & 0 & 0 & 2.40411 & 10.2987 & 69.6597 \\
 $\ep^5  \II_{3,19}$ & 0 & 0 & 0.156084 & -0.0474843 & 2.02240 & 1.77098 & 36.3488 \\
 $\ep^3  \II_{3,20}$ & 0.166667 & 1.00000 & 6.17805 & 24.1589 & 95.8265 & 291.091 & 920.800 \\
 $\ep^4  \II_{3,21}$ & 0 & -0.478563 & -2.51046 & -9.35848 & -4.69694 & 54.6072 & 447.661 \\
 $\ep^5  \II_{3,22}$ & 0 & 0 & 0 & 0 & -13.7344 & -123.498 & -834.610 \\
 $\ep^4  \II_{3,23}$ & 0 & 5.26419 & 27.6151 & 88.6126 & 76.7300 & -687.692 & -4739.22 \\
 $\ep^3  \II_{3,24}$ & 0.333333 & 1.69697 & 9.85340 & 35.3574 & 141.147 & 436.628 & 1520.17 \\
 $\ep^3  \II_{3,25}$ & -0.333333 & -0.955863 & -6.20849 & -14.1297 & -66.4814 & -137.326 & -607.205 \\
 $\ep^6  \II_{3,26}$ & 0 & 0 & 0 & 0 & 0 & 0 & -44.7364 \\
 $\ep^4  \II_{3,27}$ & 0 & -3.50946 & -13.3626 & -15.3062 & 49.5525 & 178.795 & -136.305 \\
 $\ep^6  \II_{3,28}$ & 0 & 0 & 0 & 0 & 0 & 12.6541 & 81.2200 \\
 $\ep^4  \II_{3,29}$ & 0 & 0 & -2.24115 & -12.2328 & -27.1277 & 53.4129 & 665.095 \\
 $\ep^5  \II_{3,30}$ & 0 & 0 & 0 & 0 & -13.7344 & -123.097 & -829.871 \\
 $\ep^5  \II_{3,31}$ & 0 & 0 & -2.91384 & -19.0108 & -59.3026 & -24.5645 & 686.703 \\
 $\ep^6  \II_{3,32}$ & 0 & 0 & 0 & 0 & 0 & 0 & -52.9869 \\
 $\ep^5  \II_{3,33}$ & 0 & 0 & -46.9949 & -223.713 & -517.885 & -93.8034 & -2399.21 \\
 $\ep^5  \II_{3,34}$ & 0 & 0 & 147.748 & 1759.07 & 8912.13 & 22688.0 & 23470.0 \\
 $\ep^6  \II_{3,35}$ & 0 & 0 & 0 & 0 & 0 & 19.2223 & 188.227 \\
 $\ep^5  \II_{3,36}$ & 0 & 0 & 0 & 0 & 0 & -44.3973 & -266.595 \\
 $\ep^4  \II_{3,37}$ & 0 & 0 & 0 & -131.365 & -1025.59 & -4947.66 & -17771.9 \\
 $\ep^5  \II_{3,38}$ & 0 & 0 & 0 & 0 & -13.7344 & -139.824 & -843.707 \\
 $\ep^4  \II_{3,39}$ & 0 & 5.26419 & 22.2353 & -23.3512 & -554.106 & -2700.77 & -7882.80 \\
 $\ep^3  \II_{3,40}$ & 0.333333 & 1.62253 & 9.19216 & 29.7234 & 107.761 & 264.048 & 773.366 \\
 $\ep^2  \II_{3,41}$ & -0.0833333 & -0.621542 & -4.02382 & -18.4906 & -80.1075 & -297.481 & -1082.13 \\
 $\ep^3  \II_{3,42}$ & 0.5 & 3.65481 & 17.4686 & 61.6749 & 182.326 & 462.994 & 1055.54 \\
 $\ep^3  \II_{3,43}$ & 1. & 9.75165 & 51.2036 & 189.855 & 555.345 & 1361.36 & 2906.74 \\
 $\ep^2  \II_{3,44}$ & 0.0227273 & 0.278447 & 1.87120 & 8.99092 & 34.5384 & 112.759 & 325.659 \\
 $\ep^3  \II_{3,45}$ & 0.333333 & 3.01021 & 13.5171 & 51.7380 & 143.886 & 401.974 & 841.708 \\
 $\ep^3  \II_{3,46}$ & 0.166667 & 0.977932 & 5.90734 & 21.7078 & 81.2333 & 219.743 & 637.756 \\
 $\ep^5  \II_{3,47}$ & 0 & 0 & 0 & 0 & 0 & -39.1128 & -263.481 \\
 $\ep^3  \II_{3,48}$ & 0.333333 & 1.62253 & 9.36635 & 31.5447 & 120.323 & 326.864 & 1034.42 \\
 $\ep^5  \II_{3,49}$ & 0 & 0 & 0 & 4.34479 & 39.1581 & 258.410 & 1231.82 \\
 $\ep^3  \II_{3,50}$ & 0.5 & 3.65481 & 17.6428 & 63.6870 & 195.781 & 528.695 & 1315.56 \\
 $\ep^5  \II_{3,51}$ & 0 & 0 & 0 & 4.34479 & 42.1614 & 247.132 & 1064.95 \\
\midrule 
%
%
 $\ep^5  \II_{5,1}$ & 0 & 0 & 0.333333 & 0.322119 & 4.40706 & -0.150642 & 33.9512 \\
 $\ep^5  \II_{5,2}$ & 0 & 0 & -9.04714 & -58.9589 & -215.505 & -459.906 & -429.736 \\
 $\ep^5  \II_{5,3}$ & 0 & 0 & -181.500 & -1053.68 & -3231.60 & -4909.22 & 2469.46 \\
 $\ep^5  \II_{5,4}$ & 121.000 & 701.291 & 998.063 & -6100.07 & -42256.8 & -137531. & -254864. \\
 $\ep^5  \II_{5,5}$ & 0 & 0 & 0 & -3.48876 & -11.88028 & -56.0756 & -107.5321 \\
 $\ep^5  \II_{5,6}$ & 0 & 0 & 0 & -2.11836 & -13.0865 & -52.5175 & -155.004 \\
 $\ep^5  \II_{5,7}$ & 0 & 0 & 0 & 0 & 0 & -45.0748 & -347.034 \\
 $\ep^5  \II_{5,8}$ & 0 & 0 & 0 & 19.4572 & 125.705 & 306.469 & -434.083 \\
 $\ep^6  \II_{5,9}$ & 0 & 13.4444 & 56.3818 & -126.824 & -1815.00 & -6952.38 & -9613.46 \\
 $\ep^6  \II_{5,10}$ & 0 & 0 & -168.056 & -1850.57 & -5192.00 & 16836.8 & 182559. \\
 $\ep^6  \II_{5,11}$ & 0 & -73.9444 & -787.378 & -856.028 & 17483.4 & 105004.9 & 259708. \\
 $\ep^6  \II_{5,12}$ & 0 & 0 & 0 & 0 & -5.36553 & -49.2436 & -361.801 \\
 $\ep^5  \II_{5,13}$ & -2.75 & -25.2826 & -96.8407 & -49.7955 & 1423.42 & 9014.98 & 31878.5 \\
 $\ep^6  \II_{5,14}$ & 0 & 0 & 0 & -2.20377 & -12.8942 & -34.2305 & -8.23698 \\
 $\ep^5  \II_{5,15}$ & 0 & 0 & 130.808 & 1261.41 & 5868.68 & 15139.6 & 8258.27 \\
 $\ep^5  \II_{5,16}$ & 0 & 0 & 0 & 0 & 0 & -40.5917 & -333.442 \\
 $\ep^5  \II_{5,17}$ & 0 & 0 & -2.85963 & -21.6384 & -94.8684 & -270.532 & -459.219 \\
 $\ep^5  \II_{5,18}$ & 0 & 0 & 0 & 3.48915 & 22.9870 & 138.012 & 576.781 \\
 $\ep^4  \II_{5,19}$ & 0 & 0 & -1.42155 & -12.9999 & -46.7756 & -2.92397 & 960.835 \\
 $\ep^6  \II_{5,20}$ & 0 & 0 & 0 & 0 & 0 & 17.5955 & 163.966 \\
 $\ep^5  \II_{5,21}$ & 0 & 0 & 0 & -79.3358 & -1177.69 & -10129.0 & -63928.4 \\
 $\ep^6  \II_{5,22}$ & 0 & 0 & 9.72861 & 55.7359 & 26.1826 & -1186.89 & -7813.32 \\
 $\ep^5  \II_{5,23}$ & 0 & 84.0278 & 796.920 & 1789.94 & -9909.80 & -102404. & -466731. \\
 $\ep^6  \II_{5,24}$ & 0 & 0 & 0 & 0 & 0 & -117.931 & -1692.87 \\
 $\ep^6  \II_{5,25}$ & 0 & 0 & 0 & 0 & 75.7288 & 1159.52 & 10016.2 \\
 $\ep^6  \II_{5,26}$ & 0 & 16.8056 & 62.0741 & -232.626 & -2458.85 & -9988.07 & -40585.0 \\
 $\ep^6  \II_{5,27}$ & 0 & 240.319 & 1689.06 & 6001.35 & 39657.7 & 347631. & 1.94685 10$^6$ \\
 $\ep^6  \II_{5,28}$ & 0 & -1257.06 & -9457.54 & -17640.1 & 103629.6 & 825481. & 2.91321 10$^6$ \\
\midrule 
%
%
 $\ep^5  \II_{7,1}$ & 0 & 0 & 0 & 0 & 12.8901 & 81.2922 & 481.484 \\
 $\ep^4  \II_{7,2}$ & 0 & 0 & -2.33072 & -13.8423 & -36.0947 & 19.4825 & 571.880 \\
 $\ep^5  \II_{7,3}$ & 0 & 0 & 0 & 4.47808 & 40.4644 & 268.110 & 1270.30 \\
 $\ep^4  \II_{7,4}$ & 0 & 0 & 7.38038 & 74.5639 & 389.762 & 1279.19 & 2393.58 \\
 $\ep^5  \II_{7,5}$ & 0 & 0 & 0 & 0 & 0 & -32.9413 & -359.410 \\
 $\ep^5  \II_{7,6}$ & 0 & 0 & -2.95863 & -18.1463 & -40.1238 & 38.7123 & 322.099 \\
 $\ep^6  \II_{7,7}$ & 0 & 0 & 0 & 0 & 0 & -45.2708 & -574.078 \\
 $\ep^5  \II_{7,8}$ & 0 & 0 & 43.6025 & 263.632 & 587.627 & 1275.87 & 18525.1 \\
 $\ep^5  \II_{7,9}$ & 0 & 0 & 0 & 0 & 0 & 186.457 & 2673.98 \\
 $\ep^6  \II_{7,10}$ & 0 & 0 & 0 & 0 & 0 & -124.987 & -1760.58 \\
 $\ep^6  \II_{7,11}$ & 0 & 0 & -11.0576 & -39.9400 & 9.92632 & -265.920 & -6367.54 \\
 $\ep^6  \II_{7,12}$ & 0 & 0 & 47.6893 & 471.820 & 1287.23 & 3051.84 & 55236.9 \\
 $\ep^6  \II_{7,13}$ & 0 & 0 & 73.9444 & 554.685 & 1268.98 & -181.841 & 29113.3 \\
 $\ep^5  \II_{7,14}$ & 0 & 0 & 0 & 0 & -13.7344 & -121.212 & -825.317 \\
 $\ep^4  \II_{7,15}$ & 0 & 3.50946 & 6.25551 & -40.3122 & -402.383 & -1668.35 & -3965.81 \\
 $\ep^5  \II_{7,16}$ & 0 & 0 & -2.95863 & -20.3813 & -67.7918 & -59.5060 & 582.239 \\
 $\ep^5  \II_{7,17}$ & 0 & 0 & 0.318975 & 0.575202 & 3.42490 & -0.537966 & 3.21258 \\
 $\ep^5  \II_{7,18}$ & 0 & 0 & 0 & 0 & -12.3905 & -99.5537 & -633.457 \\
 $\ep^4  \II_{7,19}$ & 0 & 1.83333 & 13.2775 & 47.4447 & 61.4885 & -312.673 & -2601.07 \\
 $\ep^6  \II_{7,20}$ & 0 & 0 & 0 & -47.7328 & -227.221 & -516.631 & -6.31579 \\
 $\ep^6  \II_{7,21}$ & 0 & 0 & 0 & -467.118 & -2321.38 & -6639.37 & -24016.0 \\
 $\ep^6  \II_{7,22}$ & 0 & 0 & 0 & 244.095 & 2124.88 & 7369.47 & 13637.2 \\
 $\ep^5  \II_{7,23}$ & 0 & 0 & -0.489073 & -2.77684 & -17.4163 & -59.8421 & -213.288 \\
 $\ep^5  \II_{7,24}$ & 0 & 0 & 0 & 0 & -12.3905 & -118.528 & -665.868 \\
 $\ep^5  \II_{7,25}$ & 0 & 0 & 8.87589 & 90.6116 & 470.191 & 1575.25 & 3474.91 \\
 $\ep^6  \II_{7,26}$ & 0 & 0 & 0 & -2.15726 & -11.1831 & -21.4410 & 60.4398 \\
\midrule 
%
%
 $\ep^6  \II_{9,1}$ & 0 & 0 & 0 & 0 & 0 & 17.5955 & 163.966 \\
 $\ep^6  \II_{9,2}$ & 0 & 0 & 0 & -4.40754 & -41.0312 & -217.292 & -804.524 \\
 $\ep^5  \II_{9,3}$ & 12.6042 & 86.7223 & 242.028 & -7.61778 & -2767.35 & -12740.9 & -36037.6 \\
 $\ep^5  \II_{9,4}$ & -2.75 & -15.8170 & -44.0632 & -40.9298 & 166.638 & 989.562 & 3008.38 \\
 $\ep^5  \II_{9,5}$ & 0 & -1.29696 & 173.737 & 849.273 & 2113.52 & 1276.27 & -9124.05 \\
 $\ep^5  \II_{9,6}$ & 0 & 9.46559 & 69.3643 & 247.676 & 438.315 & -262.731 & -4725.72 \\
 $\ep^6  \II_{9,7}$ & -6.72222 & -50.1123 & -272.224 & -1409.10 & -6197.19 & -20987.8 & -51617.2 \\
 $\ep^6  \II_{9,8}$ & 73.9444 & 847.013 & 5618.37 & 28402.0 & 120842.3 & 452504 & 1.389294 10$^6$ \\
 $\ep^6  \II_{9,9}$ & 0 & 154.821 & 1323.143 & 6232.53 & 30846.4 & 126040.2 & 413631.0 \\
 $\ep^4  \II_{9,10}$ & 0 & -0.333333 & -0.977717 & -6.34707 & -12.4124 & -58.7236 & -76.3511 \\
 $\ep^6  \II_{9,11}$ & 0 & 0 & 0 & 0 & -5.36553 & -49.2436 & -361.801 \\
 $\ep^5  \II_{9,12}$ & 0 & 0 & -9.04714 & -109.622 & -744.770 & -3340.57 & -9911.23 \\
 $\ep^6  \II_{9,13}$ & 0 & 0 & 0 & 0 & 49.9754 & 520.248 & 2808.45 \\
 $\ep^5  \II_{9,14}$ & 0 & -43.6944 & -93.3540 & 415.843 & 178.431 & -24041.1 & -181942. \\
 $\ep^6  \II_{9,15}$ & -6.72222 & -49.2349 & -201.355 & -668.570 & -1347.24 & 9095.02 & 144275.0 \\
 $\ep^6  \II_{9,16}$ & -6.72222 & 71.7651 & 1069.672 & 6620.91 & 30245.3 & 16206.1 & -557880.0 \\
 $\ep^6  \II_{9,17}$ & 73.9444 & 625.612 & 1487.16 & 11646.06 & 111041.0 & 522738.0 & 111190.3 \\
 $\ep^5  \II_{9,18}$ & 121.000 & 1112.44 & 4061.95 & 1961.06 & -52638.3 & -306082.0 & -990303.0 \\
 $\ep^6  \II_{9,19}$ & 0 & 0 & -422.272 & -3146.47 & -8236.81 & 11100.58 & 175471.0 \\
\midrule 
%
%
 $\ep^5  \II_{11,1}$ & 0 & 0 & 0 & 0 & 0 & -375.066 & -5952.68 \\
 $\ep^6  \II_{11,2}$ & 0 & 0 & 0 & 132.124 & 1491.23 & 7968.85 & 20967.9 \\
 $\ep^6  \II_{11,3}$ & 0 & -147.889 & -1803.31 & -6455.78 & -1021.87 & 136310. & 997362. \\
 $\ep^6  \II_{11,4}$ & 0 & 0 & 147.889 & 1615.38 & -2880.57 & -88377.2 & -556109. \\
 $\ep^6  \II_{11,5}$ & 0 & 0 & 0 & 0 & 0 & 0 & 337.236 \\
\midrule 
 $\ep^5  \II_{13,1}$ & 0 & 0 & -0.978147 & -8.12893 & -66.8006 & -396.511 & -2102.72 \\
 $\ep^4  \II_{13,2}$ & 0 & 0.957125 & 9.82837 & 72.3138 & 425.488 & 2148.99 & 9634.58 \\
 $\ep^5  \II_{13,3}$ & 2.52083 & 22.2984 & 92.9603 & 197.405 & -51.4506 & -2193.14 & -10204.7 \\
 $\ep^5  \II_{13,4}$ & 2.52083 & 22.2984 & 75.4773 & -52.1389 & -1803.55 & -10141.5 & -34685.7 \\
 $\ep^6  \II_{13,5}$ & 0 & 0 & 0 & 0 & 77.6668 & 1177.93 & 10104.9 \\
 $\ep^6  \II_{13,6}$ & 0 & 0 & 0 & 0 & -76.0628 & -1169.59 & -10056.1 \\
\midrule 
 $\ep^6  \II_{14,1}$ & 0 & 0 & 0 & -2.09250 & -11.7981 & -27.5242 & 26.1445 \\
 $\ep^6  \II_{14,2}$ & -6.44213 & -45.6115 & -177.925 & -583.890 & -1344.83 & 6503.66 & 119979.7 \\
 $\ep^6  \II_{14,3}$ & 6.18538 & 43.1983 & 167.975 & 589.609 & 1710.97 & -3284.57 & -99432.3 \\
\midrule 
 $\ep^5  \II_{16,1}$ & 0 & 0 & 0 & -387.864 & -3809.26 & -19780.0 & -58728.0 \\
 $\ep^5  \II_{16,2}$ & 0 & 0 & -12.0628 & -137.495 & -793.435 & -2759.34 & -4460.88 \\
 $\ep^6  \II_{16,3}$ & 0 & 0 & 0 & -3.24777 & 364.175 & 3733.95 & 19756.6 \\
\midrule 
 $\ep^6  \II_{17,1}$ & 0 & 0 & 0 & 0 & 50.9387 & 533.731 & 2918.52 \\
 $\ep^6  \II_{17,2}$ & 0 & 0 & 0 & -20.1667 & -118.669 & -1704.22 & -15275.9 \\
 $\ep^6  \II_{17,3}$ & 0 & 0 & -33.1728 & -142.466 & 855.636 & 11036.4 & 53980.0 \\
 $\ep^6  \II_{17,4}$ & 0 & 0 & -13.0062 & -213.480 & 2212.87 & 9483.34 & -119100. \\
 $\ep^6  \II_{17,5}$ & 0 & 0 & 586.735 & 5678.15 & 8686.02 & -159797. & -1.34753 10$^6$ \\
 $\ep^6  \II_{18,1}$ & 0 & 0 & 0 & -11.9014 & -134.082 & -768.979 & -2649.96 \\
 $\ep^5  \II_{18,2}$ & 0 & 0 & 0 & 4.47808 & 43.6797 & 319.065 & 1748.12 \\
 $\ep^5  \II_{18,3}$ & 0 & 0 & 8.87589 & 110.253 & 742.648 & 3324.16 & 9740.32 \\
 $\ep^6  \II_{18,4}$ & 0 & 0 & 0 & 0 & 50.9387 & 533.731 & 2918.52 \\
 $\ep^6  \II_{18,5}$ & 0 & 0 & -43.6944 & -95.1342 & 388.478 & -15.9357 & -24885.6 \\
 $\ep^6  \II_{18,6}$ & 0 & 0 & 0 & 0 & 0 & 0 & 324.818 \\
 $\ep^6  \II_{18,7}$ & 0 & 0 & -6.72222 & -43.3901 & -207.341 & -687.472 & -1073.11 \\
 $\ep^6  \II_{18,8}$ & 0 & 0 & -33.1728 & -145.736 & 808.275 & 10713.8 & 52682.6 \\
 $\ep^6  \II_{18,9}$ & 0 & 0 & 170.978 & 1816.28 & -183.561 & -96135.0 & -647509. \\
\midrule 
 $\ep^6  \II_{20,1}$ & 0 & 0 & 0 & -0.462732 & 386.584 & 3812.18 & 19786.4 \\
 $\ep^6  \II_{20,2}$ & 0 & 0 & 0 & 294.285 & 3166.91 & 6287.24 & -69674.7 \\
 $\ep^6  \II_{20,3}$ & 0 & 0 & 0 & -4.26357 & -1766.41 & -23288.6 & -143780. \\
\midrule 
 $\ep^5  \II_{21,1}$ & 0 & 0 & 0 & 0 & 0 & 176.079 & 2514.28 \\
 $\ep^5  \II_{23,1}$ & 0 & 0 & 0 & 0 & 0 & 2509.37 & 41114.2 \\
 $\ep^6  \II_{23,2}$ & 0 & 0 & 0 & 0 & 0 & 0 & -2431.37 \\
 $\ep^5  \II_{24,1}$ & 0 & 0 & 0 & 0 & 0 & 159.352 & 2083.46 \\
\bottomrule 
%
\end{longtable}
}

\subsection{Checks} 

Our results of the MIs have been checked through numerical scalar Feynman integral evaluators \texttt{FIESTA}~\cite{Smirnov:2021rhf} and \texttt{AMFlow}.
\texttt{FIESTA} can provide a few digits of precision for integrals with a smaller number of propagators. However, for integrals with a high number of propagators, such as 9, it becomes computationally very expensive and struggles to achieve even single-digit precision.
To verify our results across different domains of the kinematic variable, we employ \texttt{AMFlow} with a precision of at least 50 digits.
We evaluate the GPLs in our results using \texttt{GiNac}~\cite{Bauer:2000cp}, tabulate them, and substitute these values to obtain numerical results for the MIs at the chosen kinematic points.
Our results demonstrate perfect agreement with the \texttt{AMFlow} output across several kinematic points.

\section{Conclusion}
\label{sec:conclu}
In this paper, we have presented the three-loop master integrals with one internal massive line needed for the mixed strong-electroweak (${\mathcal{O}}(\alpha \alpha_s^2)$) corrections
to the quark form factor. 
We have employed the state-of-the-art method of differential equations to compute the master integrals. 
For topologically planar integrals, boundary values are determined through the standard approach, specifically
by either evaluating the integral at a chosen kinematic point using Feynman parameters or by imposing regularity conditions.
However, evaluating boundary values using the aforementioned procedure becomes cumbersome and impractical for more complex cases. 
To automate the determination of these constants and reconstruct their analytic expressions, we employ the auxiliary mass 
flow method in conjunction with the PSLQ algorithm.
We have encountered a few square roots in our computation, and have used variable transformations to rationalize these square roots.
However, a single transformation cannot rationalize all of them simultaneously,
and hence, we have applied different transformations for different cases. 
Consequently, we have found instances where the non-homogeneous part contains a mixture of GPLs
with two interdependent arguments, requiring each part to be integrated separately using the appropriate integration measure.
This complexity posed a significant challenge in the symbolic intermediate calculation of the MIs. 
Nevertheless, it has ultimately allowed us to obtain a concise result expressed solely in terms of GPLs, 
enabling smooth and high-precision numerical evaluation.
Given the phenomenological relevance of the DY process, our results will have an important impact in the 
physics programs at the LHC and the high-luminosity LHC. We expect to continue the evaluations of
the associated form factors in future work.

\section*{Acknowledgment}
We would like to thank S. Moch, V. Ravindran, A. Saha and A. Vicini for fruitful discussions and their comments on the manuscript.
N.R. sincerely thanks the University of Hamburg and CERN for their hospitality during the finalization of this work.
N.R. is partially supported by the SERB-SRG under Grant No. SRG/2023/000591.

\bibliography{main} 
\bibliographystyle{JHEP}

\end{document}